\documentclass[twocolumn,aps,prd,groupedaddress,nofootinbib,showpacs]{revtex4-1}
\usepackage{graphicx}
\usepackage{amsmath}
\usepackage{bm}
\usepackage{slashed}
\usepackage{epsfig}
\usepackage{amsfonts}
\usepackage{color}
\usepackage{epstopdf}
\usepackage{enumerate}
\usepackage{hyperref}

\newcommand{\state}[4]{{^{#1}\hspace{-0.6mm}#2_{#3}^{[#4]}}}

\newcommand\COcSa{\state{3}{S}{1}{8}}

\newcommand{\onia}{{H}}
\newcommand{\qq}{Q{\overline{Q}}}

\begin{document}

\title{New factorization theory for heavy quarkonium production and decay}


\author{Yan-Qing Ma$^{1,2,3}$}
\author{Kuang-Ta Chao$^{1,2,3}$}
\affiliation{
$^{1}$School of Physics and State Key Laboratory of Nuclear Physics and
Technology, Peking University, Beijing 100871, China\\
$^{2}$Center for High Energy Physics,
Peking University, Beijing 100871, China\\
$^{3}$Collaborative Innovation Center of Quantum Matter,
Beijing 100871, China
}%
\date{\today}

\begin{abstract}
The widely used  nonrelativistic QCD (NRQCD) factorization theory now encounters some notable difficulties in describing quarkonium production.  This may be due to the inadequate treatment of soft hadrons emitted in the hadronization process, which causes bad convergence of velocity expansion in NRQCD. In this paper, starting from QCD we propose a rigorously defined factorization approach, soft gluon factorization (SGF), to better deal with the effects of soft hadrons. After a careful velocity expansion, the SGF can be as simple as the NRQCD factorization in phenomenological studies, but has a much better convergence. The SGF may provide a new insight to understand the mechanisms of quarkonium production and decay.
\end{abstract}


\maketitle

\section{Introduction}

As the simplest bound state of strong interactions, heavy quarkonium is a perfect system to study both perturbative and nonperturbative physics of QCD. Ever since the discovery of the first heavy quarkonium $J/\psi$ in 1974, a lot of efforts have been devoted to interpret the production and decay mechanisms of heavy quarkonium. Among them, the most notable theories include the color-singlet model \cite{Ellis:1976fj,Carlson:1976cd,Chang:1979nn} and the nonrelativistic QCD (NRQCD) factorization theory \cite{Bodwin:1994jh} \footnote{For quarkonium production only, there are two more well-established theories. One is the color-evaporation model \cite{Fritzsch:1977ay,Halzen:1977rs,Ma:2016exq}, which is a phenomenological model that may not be able to derive from first principles of QCD.  The other is QCD collinear factorization \cite{Kang:2011mg,Fleming:2012wy,Kang:2014tta}, which is rigorous but can only describe high momentum quarkonium production.}.

The NRQCD factorization approach is successful. With the color-octet mechanism, NRQCD can solve the infrared divergence problem encountered in the color-singlet model \cite{Bodwin:1992qr},
explain the $\psi(2S)$ surplus \cite{Braaten:1994vv}, and describe the inclusive quarkonium production \cite{Boyd:1998km,Kramer:2001hh,Klasen:2001cu,Zhang:2006ay,Lansberg:2008gk,Ma:2008gq,Gong:2009kp,Butenschoen:2009zy,Ma:2010vd,Ma:2010yw,Butenschoen:2010rq,Ma:2014mri}. In addition, although there is still no convincing all-order proof of NRQCD factorization for quarkonium production, it was found that factorization may hold at least to next-to-next-to-leading order (NNLO) if long-distance matrix elements (LDMEs) are modified to be gauge complete \cite{Nayak:2005rt,Nayak:2005rw,Nayak:2006fm}.

Nevertheless, studies in recent years have shown that NRQCD factorization in describing quarkonium production may encounter some notable difficulties.
(1) {The polarization puzzle:} The leading order calculation in NRQCD implies that $\psi(nS)$ and $\Upsilon(nS)$ produced at hadron colliders are transversely polarized due to $\COcSa$ channel dominance \cite{Cho:1994ih,Beneke:1995yb,Braaten:1999qk}. But experimental measurements found these states almost unpolarized \cite{Abulencia:2007us,Abelev:2011md,Aaij:2013nlm,Chatrchyan:2013cla,Chatrchyan:2012woa}. Thanks to next-to-leading order calculations \cite{Gong:2008hk,Butenschoen:2012px,Chao:2012iv,Gong:2012ug,Bodwin:2014gia,Faccioli:2014cqa,Gong:2013qka,Han:2014kxa}, the observed polarizations of $J/\psi$ and $\Upsilon(nS)$ can be qualitatively explained, but it is still hard to understand the polarization of $\psi(2S)$ \cite{Shao:2014yta}.
(2) {The hierarchy problem:} The best fit of $J/\psi$ yield data at high transverse momentum in hadronic collisions determines two linearly combined LDMEs, $M_0=0.074\text{GeV}^3$ and $M_1=0.0005\text{GeV}^3$ \cite{Ma:2010yw} (the $J/\psi$ polarization data requires almost the same two combined LDMEs \cite{Chao:2012iv}). There is a two-orders difference between the two combined LDMEs.  However, velocity scaling rules in NRQCD  \cite{Bodwin:1994jh} expect these LDMEs to be at the same order of magnitude.
(3) {The universality problem:} A necessary condition for NRQCD factorization is that LDMEs are universal, i.e. process independent. Yet phenomenology studies show that $M_0$ extracted from hadron colliders \cite{Ma:2010yw,Butenschoen:2010rq,Gong:2012ug} is much larger than the upper bound set by $e^+ e^-$ collisions $M_0<0.02\text{GeV}^3$ \cite{Zhang:2009ym}.

In this paper, we show that the velocity expansion in the present NRQCD framework suffers from large high order relativistic corrections due to ignoring the momentum of soft hadrons, which are mainly soft gluons perturbatively, emitted in the hadronization process. Thus, by including only a few low order contributions in relativistic expansion, NRQCD is hard to provide good descriptions for quarkonium production, which may be the reason for the above mentioned difficulties.
Starting from QCD, we propose a new factorization approach, called soft gluon factorization (SGF), to better deal with the effects of soft hadrons. The SGF has a much better convergence in the velocity expansion,
and it may provide a new insight to fully understand the mechanisms of quarkonium production and decay.

The rest of the paper is organized as follows. In Sec. \ref{sec:fac}, we explain why NRQCD factorization may have bad convergence in velocity expansion, and then define the SGF formula with rigorous operator definition. Relations to other approaches are also discussed. In Sec.~\ref{sec:sim}, we do a careful simplification to the SGF so that it will be easy to use in practice, but at the same time it can capture main physics. Following this, we compare the SGF and NRQCD by studying the gluon fragmentation to yield of $J/\psi$ in Sec.~\ref{sec:jpsi}. As expected, we find that the lowest order NRQCD approximation is not good, which may be the reason why NRQCD encounters many difficulties. The summary and the possibility of solving these difficulties in the SGF framework are given in Sec.~\ref{sec:sum}.

\section{Soft gluon factorization}
\label{sec:fac}

\subsection{Bad convergence of $v^2$ expansion in NRQCD}\label{sec:badconv}

In NRQCD factorization \cite{Bodwin:1994jh}, the differential cross section of a heavy quarkonium $\onia$ production can be factorized as
\begin{align}\label{eq:facNR}
\begin{split}
(2\pi)^3 2 P_\onia^0 \frac{d\sigma_\onia}{d^3P_\onia}=& \sum_{n} {\cal H}_{n}(P_\onia)\, \langle {\cal O}_n^\onia \rangle+\cdots,
\end{split}
\end{align}
where $P_\onia$ is the momentum of $\onia$, $n$ denotes intermediate $Q{\overline{Q}}$ states, whose quantum numbers are usually expressed in terms of spectroscopic notation $\state{{2S+1}}{L}{J}{c}$, with $c=1, 8$ denotes color singlet or color octet of the pair, ${\cal H}_{n}(P_\onia)$ are perturbative calculable short-distance coefficients which can be expanded order by order in $\alpha_s$, $ \langle {\cal O}_n^\onia \rangle$ are gauge-completed \cite{Nayak:2005rt,Nayak:2005rw,Nayak:2006fm} nonperturbative long-distance matrix elements (LDMEs) which can be classified according to the power counting rules of $v$, and the ellipses denote other relativistic correction terms. It is needed to emphasize that, in NRQCD factorization, one expands the mass of quarkonium  $M_\onia$ around twice of heavy quark mass $2m$, which results in $P_\onia^2\approx4m^2$ in Eq.~\eqref{eq:facNR}.

If convergence of velocity expansion in NRQCD factorization is very good, one only needs to consider a very limited number of LDMEs to describe experimental data. Unfortunately, it may not be the case. To see this, for an example let us consider the differential cross section of quarkonium production in hadron colliders with high transverse momentum $P_{\onia T}$. In NRQCD factorization, one calculates the partonic $\qq$ production differential cross sections with transverse momentum $P_{T}$, and then expands $P_T$ around $P_{\onia T}$. On average, let us set $P_{\onia T}\sim (1-\lambda)P_T$, with $\lambda$ being the average transverse momentum fraction carried by soft gluons which can be at the order of $v$ or $v^2$. At a specific range of large $P_{\onia T}$, differential cross section behaves as
\begin{align}\label{eq:power}
\begin{split}
\frac{d\sigma_\onia}{dP_{\onia T}^2}&\sim \frac{1}{P_{T}^c}\sim \frac{(1-\lambda)^c}{ P_{\onia T}^c},
\end{split}
\end{align}
with $c$ being usually larger than 4. Expansion of this result with respect to $\lambda$ mimics the relativistic expansion in NRQCD. For a typical choice $\lambda=0.2$ and $c=4$~\cite{Mangano:1996kg}, we have $(1-\lambda)^c=0.4096$, but the expansion of $(1-\lambda)^c$ to $O(\lambda^0)$ gives 1, and expansion to $O(\lambda^1)$ gives 0.2. The above argument shows a bad convergence of relativistic expansion. We note that the convergence still cannot be improved if one instead is to calculate, e.g., $P_{\onia T}^4\frac{d\sigma_\onia}{dP_{\onia T}^2}$. Furthermore, as we will see in Sec.~\ref{sec:jpsi}, there are other relativistic correction terms that make the convergence of relativistic expansion even worse.

Because of the bad convergence of relativistic corrections, theoretic calculations based on only a limited number of LDMEs are sometimes hard to describe experimental data. In literature, e.g. Refs.~\cite{Beneke:1997qw,Fleming:2003gt,Fleming:2006cd,Leibovich:2007vr}, for each problem one can resum a specific subset of LDMEs to improve the theoretic results. The resummation results in many so-called ``shape functions".

A factorization method aiming to improve convergence for an arbitrary problem was proposed in Ref. \cite{Beneke:1999gq}. However, this method can only be thought of as a model because there is no operator definition for nonperturbative functions. Without an operator definition, the method is not well defined and it is not possible to do rigorous calculation beyond tree level.

\subsection{Soft gluon factorization formula}

The aim of soft gluon factorization is to resum a subset of relativistic correction terms in NRQCD factorization that are important for a phenomenological purpose. It is convenient to demand the subset to be Lorentz invariant. The formula of SGF for a quarkonium $\onia$ production is
\begin{align}\label{eq:fac4d}
 (2\pi)^3 2 P_\onia^0 \frac{d\sigma_\onia}{d^3P_\onia}\approx \sum_{n} \int \frac{d^4 P}{(2\pi)^4} {\cal H}_{n}(P) F_{n\to \onia}(P,P_\onia),
\end{align}
where ${\cal H}_{n}(P)$ are perturbatively calculable hard parts that, roughly speaking, produce an intermediate state with quantum numbers $n$ and momentum $P$, and $F_{n\to \onia}(P,P_\onia)$ are nonperturbative functions, which we call soft gluon distribution functions (SGDs), which describe the hadronization of the intermediate state to physical quarkonium $\onia$. To account for the effect of soft hadrons emission, which are mainly soft gluons perturbatively,  the momentum of the observed quarkonium $P_\onia$ is kept different from the momentum of the intermediate state $P$, which is different from the treatment in NRQCD.

For quarkonium decay, one can define a similar formula.

\subsection{Intermediate states}

Naively, in Eq.~\eqref{eq:fac4d} we should sum over a complete set of intermediate states that may or may not contain a $Q{\overline{Q}}$ pair. However, if there is no $Q{\overline{Q}}$ pair in the intermediate state, a $Q{\overline{Q}}$ pair must be produced in the hadronization process. In this case, the corresponding $F_{n\to \onia}(P,P_\onia)$ has short-distance effects and can be refactorized, which enables us to eliminate this contribution from the factorization formula. Furthermore, if the intermediate state contains other energetic partons in addition to a $Q{\overline{Q}}$ pair, the corresponding $F_{n\to \onia}(P,P_\onia)$ can also be refactorized because there is no energetic gluon or light quark in the dominant Fock state of conventional quarkonium.  In this consideration, we only need to sum over intermediate states which contain a $Q{\overline{Q}}$ pair and some soft partons. According to physical scales existing in the quarkonium system, energy of these soft partons can be at the order of $m v$ or $m v^2$, which are assumed to be much smaller than $m$.

As each conventional quarkonium has a $Q{\overline{Q}}$ state with specific quantum numbers as its leading Fock state, the $Q{\overline{Q}}$ intermediate state with the same quantum numbers has leading contribution. If the intermediate state contains soft partons in addition to a $Q{\overline{Q}}$ pair, which means that soft partons are produced in the hard parts, the contribution is at least suppressed by $v^2$ comparing with the contribution from the intermediate state with the same $Q{\overline{Q}}$ but without soft partons. In other words, power counting of the corresponding $F_{n\to \onia}(P,P_\onia)$ in the factorization formula is at high order in $v^2$. Unlike $v^2$ corrections due to kinematic effects of soft gluons emission, contribution of intermediate states with soft partons should have a mild $v^2$ corrections. Thus, as the first approximation, we will ignore these contributions and consider only intermediate states constituted by a $Q{\overline{Q}}$ pair in the rest of the paper.

Then $n$ in Eq.~\eqref{eq:fac4d}  denotes intermediate $Q{\overline{Q}}$ states, whose quantum numbers are usually expressed in terms of spectroscopic notation $n=Q{\overline{Q}}(\state{{2S+1}}{L}{J,J_z}{c})$, with $c=1, 8$ denoting color singlet or color octet of the pair. We note that the state in the amplitude $n$ can be in general different from that in the complex-conjugate amplitude $\tilde{n}=Q{\overline{Q}}(\state{{2\tilde{S}+1}}{\tilde{L}}{\tilde{J},\tilde{J}_z}{\tilde{c}})$.  Color charge, $C$-parity, $P$-parity and angular momentum conservations provide general selection rules \cite{Ma:2015yka}, which gives that the following relations always hold: $c=\tilde{c}$, $S=\tilde{S}$, and $|L-\tilde{L}|=0,2,4,\cdots$. For the production of a polarization-summed quarkonium, we have further constraints $J=\tilde{J}$, $J_z=\tilde{J}_z$. Thus, even in the polarization-summed case, there is interference between the $\state{{3}}{S}{1,J_z}{c}$ state and the $\state{{3}}{D}{1,J_z}{c}$ state. Selection rules for polarization distributions can be found in Ref.~\cite{Ma:2015yka}.
For simplicity, in the rest of the paper we will only discuss polarization-summed quarkonium production and choose the quantum numbers of the intermediate $Q{\overline{Q}}$ pair in the amplitude to be the same as that in the complex-conjugate amplitude, while polarization and interference contributions can be studied similarly.

\subsection{Nonperturbative distributions}

To make SGDs Lorentz invariant, it is convenient to define them using QCD fields instead of NRQCD fields. We define SGDs for polarization-summed quarkonium $\onia$ production as vacuum expectation values of bilocal operators constructed from QCD fields,
\begin{align}\label{eq:SGDs}
	\begin{split}
		&F_{n\to \onia}(P,P_\onia)=\int d^4b e^{-iP\cdot b}\,\\
		&\times \langle 0| [\overline\Psi{\cal K}_{{n}}\Psi]^{\dagger}(0) \big(a_\onia^\dagger a_\onia\big) [\overline\Psi {\cal K}_n\Psi](b) |0\rangle_{\text{S}},
	\end{split}
\end{align}
where
\begin{align}
	a_\onia^\dag a_\onia= \sum_X\sum_{J_z^\onia} |\onia+X\rangle \langle \onia+X|
\end{align}
projects final state to include a polarization-summed $\onia$ with relativistic normalization $\langle\onia(P_\onia)|\onia(P_\onia')\rangle=(2E_\onia)(2\pi)^3\delta^3({\bm P}_\onia-{\bm P}_\onia')$. The subscript ``S" means that, to evaluate the matrix element, one only picks up integration regions where off-shellness of all particles is much smaller than heavy quark mass \footnote{From the point view of method of region \cite{Beneke:1997zp}, the effect of ``S" keeps all regions except the hard region.
}.  ${\cal K}_n(r b)$ are projection operators defining the intermediate state $n$, 
\begin{align}\label{eq:Kn}
	{\cal K}_n(r b)=\frac{\sqrt{M_\onia}}{M_\onia+2m}\frac{M_\onia+\slashed{P}_\onia}{2M_\onia}\Gamma_{n}\frac{M_\onia-\slashed{P}_\onia}{2M_\onia}\,  {\cal C}^{[c]}\,,
\end{align}
where angular momentum operators $\Gamma_{n}$ are defined explicitly in Appendix \ref{sec:defangular}, color operators ${\cal C}^{[c]}$ will be defined later, $\frac{M_\onia\pm\slashed{P}_\onia}{2M_\onia}$ project out large components in velocity expansion, and, as we will see later, the introduction of prefactor $\frac{\sqrt{M_\onia}}{M_\onia+2m}$ makes our SGDs more closely related to NRQCD LDMEs.

If the intermediate $\qq$ pair is in color singlet, we define color operator as
\begin{align}
		{\cal C}^{[1]}&=\frac{{\bm 1}_c}{\sqrt{N_c}},
\end{align}
which is nothing but SU(3) color Clebsch-Gordan coefficient $\langle 3i;\bar{3}\bar{i}|00 \rangle$ and ${\bm 1}_c$ is the identity matrix in color space.
If the intermediate $\qq$ pair is in color octet, we define color operator by the multiplication of a color matrix in fundamental representation with a gauge link in adjoint representation,
\begin{align}
		{\cal C}^{[8]}&=\sqrt{2}t^{\bar{a}}\, \Phi^{(A)}_{a\bar{a}}(r b),
\end{align}
where $\sqrt{2}t^{\bar{a}}$ is SU(3) color Clebsch-Gordan coefficient $\langle 3i;\bar{3}\bar{i}|8\bar{a} \rangle$ and the introduction of gauge link $\Phi^{(A)}_{\bar{a}a}(r b)$ is to enable gauge invariance of SGDs.

The next question is what direction of the gauge link should we choose. A natural choice is to define the direction along a light cone, because a gauge link can be obtained by eikonal approximation for soft gluon interaction between the $Q{\overline{Q}}$ pair and jets. Indeed, this is the choice to define gauge-completed NRQCD LDMEs in Refs.~\cite{Nayak:2005rt,Nayak:2005rw,Nayak:2006fm}. In the SGF framework, however, this choice will result in uncanceled gauge-link-collinear rapidity divergence because the position $b$ in Eq.~\eqref{eq:SGDs} is in general not along a light cone. This is the same problem encountered in transverse-momentum-dependent (TMD) factorization where there is a mismatch of rapidity divergence between virtual correction and real correction \cite{Collins:2011zzd}. To cancel the rapidity divergence, soft factor as while as jet functions may need to be introduced.

If we do not want to introduce additional soft factors \footnote{As we will discuss later, the SGF Eq.~\eqref{eq:fac4d} is only a conjecture and it is not clear right now whether it is valid to all orders. If Eq.~\eqref{eq:fac4d} does not hold to all orders, maybe a modification by introducing a gauge invariant soft factor is unavoidable.}, the gauge links can be defined
along the $b$ direction,
\begin{align}\label{eq:glk}
\Phi^{(A)}(r b)={\cal P}\,{\text{exp}}\left\{ -ig_s \int_0^\infty d\lambda\, b_\ell\cdot A^{(A)}(r\, b+\lambda\,b_\ell)\right\},
\end{align}
where ${\cal P}$ denotes path ordering, $A^{(A)}$ is gluon field with color matrix in adjoint representation, and $b_\ell^\mu=b^\mu+\varepsilon \ell^\mu$. We choose $0<\varepsilon\ll1$ and a fixed light-like vector $\ell^\mu$ so that: when $b^\mu$ is finite, $b_\ell^\mu$ is the same as $b^\mu$; while as $b^\mu\to0$, the gauge link is well defined (along the $\ell^\mu$ direction). 

In the above, we have defined SGDs for $n$ with fixed $J_z$. SGDs for polarization-summed intermediate states can be defined accordingly,
\begin{align}
	\begin{split}
		&F_{Q{\overline{Q}}(\state{{2S+1}}{L}{J}{c})\to \onia}\equiv\sum_{J_z} F_{Q{\overline{Q}}(\state{{2S+1}}{L}{J,J_z}{c})\to \onia}\,.
	\end{split}
\end{align}

Similar to LDMEs in NRQCD, each SGD has a definite power counting in $v^2$, which will be explained later. Thus, for a certain accuracy,  we can truncate to use only a few SGDs for each $\onia$ production.

\subsection{Perturbative matching}

With the definition of SGDs, short-distance hard parts can be obtained order by order in perturbation theory by matching both sides of Eq.~\eqref{eq:fac4d}. As ${\cal H}_{n}(P)$ are independent of nonperturbative physics, they are unchanged if we replace the quarkonium $\onia$ in Eq.~\eqref{eq:fac4d} by an on-shell $Q{\overline{Q}}$ pair with specific quantum numbers. To make the following equations more convenient to use, we first relabel $n$ and $P$ in Eq.~\eqref{eq:fac4d} by $n'$ and $P'$, respectively, and we then replace $\onia$  in both sides of the equation by a state $n=Q{\overline{Q}}(\state{{2S+1}}{{L}}{{J,J_z}}{{c}})$ with momentum $P$, which results in 
\begin{align} \label{eq:matching}
\overline{d\sigma}_{n}(P)\approx \sum_{n'} \int \frac{d^4 P'}{(2\pi)^4} {\cal H}_{n'}(P') F_{n'\to n}(P',P)\,,
\end{align}
where
\begin{align}
\overline{d\sigma}_{n}(P)\equiv(2\pi)^3 2 P^{0} \frac{d\sigma_{n}}{d^3P}.
\end{align}
Note that, although $\overline{d\sigma_\onia}$ and $F_{n'\to \onia}$ are changed to  $\overline{d\sigma_{n}}$ and $F_{n'\to n}$, respectively, ${\cal H}_{n'}(P')$ stays unchanged, which is guaranteed by the fact that ${\cal H}_{n'}(P')$ is perturbatively calculable and thus it is independent of nonperturbative external states. The reason for choosing the on-shell $Q{\overline{Q}}$ pair is that $\overline{d\sigma_{n}}$ and $F_{n'\to n}$ are then gauge invariant and can be expanded order by order in $\alpha_s$ in any gauge.

The key to determine all ${\cal H}_{n'}$ is to project $\onia$ to a complete set of $Q{\overline{Q}}$  states, and then to match coefficients of $\alpha_s$ in both sides of Eq.~\eqref{eq:fac4d}. Although any complete set is doable for this purpose, a good choice should satisfy as much as possible that leading order in $\alpha_s$ expansions of $F_{n'\to n}$ are delta functions.

To define a good complete set, we
begin with a  $Q{\overline{Q}}$ with momenta
\begin{align}
p_Q=P/2+q, \hspace{1cm} p_{{\overline{Q}}}=P/2-q\,.
\end{align}
We project it to color state $c$ by color projectors:
\begin{subequations}
	\begin{align}
	&\text{$\frac{{\bm 1}_c}{\sqrt{N_c}}$ for color singlet},\\
    &\text{$\sqrt\frac{2}{N_c^2-1}t^{a}$ for color octet},
	\end{align}
\end{subequations}
where $\sqrt{N_c^2-1}$ in the denominator is to average over color-octet states. We then project it to state with total spin $S$ and $S_z$ by replacing spinors of $Q{\overline{Q}}$ by\footnote{This expression can be interpreted as covariant spin projectors \cite{Kuhn:1979bb,Guberina:1980dc} multiplied by a normalization factor $\sqrt{2/M}$ \cite{Fan:2009cj}. But for our purpose, this interpretation is unnecessary. The key in our definition is to guarantee the orthogonal relations Eq.~\eqref{eq:ortho}.}
\begin{align}
	\frac{2}{\sqrt{M}(M+2m)}(\slashed{p}_{\overline Q}-m) \frac{M-\slashed{P}}{2M} \widetilde{\Gamma}^s_{SS_z} \frac{M+\slashed{P}}{2M} (\slashed{p}_{ Q}+m),
\end{align}
where $M^{2}=P^{2}$ and
\begin{subequations}\label{eq:spinProjBar}
	\begin{align}
	\widetilde{\Gamma}^s_{00} &= - \gamma_5,\\
	\widetilde{\Gamma}^s_{1S_z} &= {\epsilon}_{S_z}^{*\mu} \gamma_\mu,
	\end{align}
\end{subequations}
On-shell conditions $p_Q^{ 2}=p_{{\overline{Q}}}^{2}=m^2$ result in
\begin{align}\label{eq:onshell}
P\cdot q=0,\hspace{1cm} q^{ 2}=m^2-P^{2}/4\,.
\end{align}
There are still 2 degrees of freedom of the relative momentum $q$, which can be chosen as spatial angles in the rest frame of the pair. We further do a partial wave expansion in this frame, which removes the $q$ dependence although introducing orbital angular momentum $L$ and $L_z$. Finally, we project spin and orbital angular momentum to total momentum $J$ and $J_z$. More precisely, in the rest frame of $P$, $d\sigma_{n}$ and $F_{n'\to n}$ with quantum number $n=Q{\overline{Q}}(\state{{2S+1}}{{L}}{{J,J_z}}{{c}})$ can be obtained from the corresponding production of $\qq$ with fixed $c$, $S$ and $S_z$ by the following operation:
\begin{align}
\sum_{L_z,S_z} \langle L, L_z; S,S_z |J,J_z \rangle \int d^2 \Omega  |{\bm q}|^{-L} \sqrt{\frac{(2L+1)!!}{4\pi (L!)}} Y_{L}^{*L_z}\,.
\end{align}
In Appendix \ref{sec:checkLO}, we show that the above definition can indeed result in orthogonal relations at lowest order
\begin{align}\label{eq:ortho}
F^{(0)}_{n'\to n}(P',P)=\delta_{n' n} (2\pi)^4 \delta^4(P'-P)\,,
\end{align}
where the superscript ``$(0)$'' denotes leading order in $\alpha_s$ expansion.

By inserting perturbative expansions
\begin{subequations}
	\begin{align}
	\overline{d\sigma}_{n}&=\overline{d\sigma}_{n}^{(0)}+\alpha_s \overline{d\sigma}_{n}^{(1)}+\alpha_s^2 \overline{d\sigma}_{n}^{(2)}+\cdots,\\
	F_{n'\to n}&=F_{n'\to n}^{(0)}+\alpha_s F_{n'\to n}^{(1)}+\alpha_s^2 F_{n'\to n}^{(2)}+\cdots,\\
	{\cal H}_{n}&={\cal H}_{n}^{(0)}+\alpha_s {\cal H}_{n}^{(1)}+\alpha_s^2 {\cal H}_{n}^{(2)}+\cdots
	\end{align}
\end{subequations}
into Eq.~\eqref{eq:matching} and using the orthogonal relations Eq.~\eqref{eq:ortho}, coefficients of different powers of $\alpha_s$ give the following relations:
\begin{subequations}\label{eq:match}
	\begin{align}
{\cal H}_{n}^{(0)}(P)=&\overline{d\sigma}_{n}^{(0)}(P),\\
\begin{split}
{\cal H}_{n}^{(1)}(P)=&\overline{d\sigma}_{n}^{(1)}(P)\\
&\hspace{-2mm}-\sum_{n^\prime} \int \frac{d^4 P^\prime}{(2\pi)^4} \overline{d\sigma}_{n^\prime}^{(0)}(P^\prime) F_{n^\prime\to n}^{(1)}(P^\prime,P),
\end{split}\\
\begin{split}
{\cal H}_{n}^{(2)}(P)=&\overline{d\sigma}_{n}^{(2)}(P)\\
&\hspace{-2mm}-\sum_{n^\prime} \int \frac{d^4 P^\prime}{(2\pi)^4} \overline{d\sigma}_{n^\prime}^{(1)}(P^\prime) F_{n^\prime\to n}^{(1)}(P^\prime,P)\\
&\hspace{-2mm}-\sum_{n^\prime} \int \frac{d^4 P^\prime}{(2\pi)^4} \overline{d\sigma}_{n^\prime}^{(0)}(P^\prime) F_{n^\prime\to n}^{(2)}(P^\prime,P),
\end{split}
\end{align}
\end{subequations}
and so on. Based on these relations, ${\cal H}_{n}$ can be obtained by perturbative calculation of $\overline{d\sigma}_{n}$ and $ F_{n^\prime\to n}$ for on-shell $Q{\overline{Q}}$ pair production. This perturbative calculation relies on the cancellation of IR divergences between $\overline{d\sigma}_{n}$ and $ F_{n^\prime\to n}$ [like in Eq.~\eqref{eq:match}] to all orders in perturbation theory, which is a very difficult problem and will be discussed in the next subsection.

It is worth emphasizing that the above perturbative calculation of on-shell $Q{\overline{Q}}$ pair production can only directly determine  ${\cal H}_{n}(P)$ with $P^2>4m^2$. Although on average we may have $P^2>4m^2$ \cite{Ma:2016exq}, there is a nonvanishing contribution from the $P^2<4m^2$ region in Eq.~\eqref{eq:fac4d}. The value of ${\cal H}_{n}(P)$ in the later region can be obtained by analytical continuation of its value in the former region. Analytical continuation here is almost trivial, which means that we use the same functional form of ${\cal H}_{n}(P)$ in all regions of $P^2$.

\subsection{Justification}

The above proposed SGF is only a conjecture, and we cannot provide an all order proof at present. The most dangerous interaction that may ruin the factorization is the elastic scattering between the $Q\bar Q$ pair and hard jets by exchanging gluons, the kinematic region of these exchanged gluons are called ``Glauber" region in literature. We find that IR divergence from the ``Glauber'' region at one-loop order is purely imaginary, which cancels with the contribution from its complex conjugate diagram. Therefore, SGF holds at one-loop order. Thanks to the two-loop study of infrared divergences of quarkonium production~\cite{Nayak:2005rt,Nayak:2005rw,Nayak:2006fm}, it is hopeful that SGF may also hold to two-loop order, which however still needs further examination.

\subsection{Relation to NRQCD factorization}

The difference between NRQCD and SGF can be partially understood from the treatment of intermediate momenta. Based on relations in Eq.~\eqref{eq:onshell}, we can express $q'_0$ and $\bm q'^2$ by $P'_0$ and $\bm P'^2$, which corresponds to the SGF strategy. Alternatively, we can also use Eq.~\eqref{eq:onshell} to express $q'_0$ and $P'_0$ by $\bm q'^2$ and $\bm P'^2$, which corresponds to the strategy of NRQCD factorization.

To further discuss the relation between SGF and NRQCD, by ignoring velocity corrections we can approximate ${\cal H}_{n}(P)$  by ${\cal H}_{n}(P_\onia)$ in Eq.~\eqref{eq:fac4d}. Then the integral over $P$ is applied on SGDs defined in Eq.~\eqref{eq:SGDs}, and thus Eq.~\eqref{eq:fac4d} becomes
\begin{align}
 (2\pi)^3 2 P_\onia^0 \frac{d\sigma_\onia}{d^3P_\onia}\approx \sum_{n} {\cal H}_{n}(P_\onia) \langle \widetilde{\mathcal O}_n^\onia \rangle,
\end{align}
which has the same form as the NRQCD factorization Eq.\eqref{eq:facNR}. In the above formula,
\begin{align}
\langle \widetilde{\mathcal O}_n^\onia \rangle=\int \frac{d^4P}{(2\pi)^4} F_{n\to\onia}(P,P_\onia)
\end{align}
are closely related to gauge-completed LDMEs $\langle {\mathcal O}_n^\onia \rangle$ although they are defined by different fields. Considering the differences between normalization in color space, up to lowest order in $v^2$ approximation, we have \footnote{Let us first check color-octet states. In NRQCD one usually chooses nonrelativistic normalization for the $\onia$ state, which means that our definition has an extra factor $2E_\onia$ due to normalization of state. For a color operator, in NRQCD one uses $t^{\bar{a}}$ instead of $\sqrt{2}t^{\bar{a}}$, due to which we have an extra factor $2$. Finally, the prefactor in Eq.~\eqref{eq:Kn} introduces an extra factor $M_\onia/(M_\onia+2m)^2$. The multiplication of these three factors equals $1$ if we choose the $\onia$ rest frame and take $v^2\to0$. Except these differences, all other parts of our definition are identical to the NRQCD definition at the lowest order in $v^2$ approximation. For color-singlet states, in NRQCD one uses $\bm{1}_c$ as color operator, which results in a difference of factor $2N_c$ comparing with the color-octet case. }
\begin{subequations}\label{eq:ldmes}
\begin{align}
\langle {\mathcal O}_n^\onia \rangle\approx & 2N_c \langle \widetilde{\mathcal O}_n^\onia \rangle,~~~\text{if $n$ is color singlet},\\
\langle {\mathcal O}_n^\onia \rangle\approx & \langle \widetilde{\mathcal O}_n^\onia \rangle,~~~\text{if $n$ is color octet}.
\end{align}
\end{subequations}
Based on NRQCD velocity scaling rules, these relations also tell us $v^2$ power counting rules for SGDs.

Beyond the above approximation, the complete SGF formula resums a series of velocity corrections in NRQCD, which are important for phenomenological study. As we know, the TMD factorization is a generalization of collinear factorization with a series of higher twist corrections resummed  \cite{Collins:2011zzd}. In this sense, we can say that SGF is a ``TMD" version of NRQCD. We expect that there are a lot of similarities between SGF and TMD factorization.

\subsection{Relation to QCD collinear factorization}

It is also needed to explain the relation between the SGF method and the QCD collinear factorization method \cite{Kang:2011mg,Fleming:2012wy,Kang:2014tta}. In QCD collinear factorization, one assumes the transverse momentum of the produced quarkonium to be much larger than the heavy quark mass, $P_{HT}\gg m$. Therefore, one can factorize out hard physics at the scale of $P_{HT}$, and leave the physics at the scale of $m$ to be described by input functions, which are called fragmentation functions (FFs). Unlike NRQCD factorization and SGF, QCD collinear factorization has been proved to all orders in perturbation theory at both leading power and next-to-leading power levels. Phenomenologically, the advantage of this method is that large logarithms $\log(P_{HT}/m)$ can be resummed to all orders in perturbation theory using renormalization group equations for single-parton and double-parton FFs \cite{Kang:2014tta}, and therefore there can be smaller theoretical uncertainties.

Due to the above advantages, when studying high transverse momentum quarkonium production, one should better first use QCD collinear factorization to express cross sections in terms of the convolution of perturbative calculable hard parts \cite{Kang:2014pya} with FFs. Based on renormalization group equations, these FFs are fully determined by their values at an initial scale. Because of the heavy quark mass, these FFs at the initial scale still have perturbative physics that can be further separated from nonperturbative physics. This can be achieved by using NRQCD or SGF. Using NRQCD factorization, these FFs at an initial scale have been studied extensively \cite{Braaten:1993mp,Braaten:1993rw,Braaten:1994kd,Braaten:1995cj,Ma:1995vi,Braaten:2000pc,Bodwin:2003wh,Hao:2009fa,Jia:2012qx,Bodwin:2012xc,Ma:2013yla,Ma:2014eja}. In Sec. \ref{sec:jpsi}, we will see that FFs can also be calculated by using the SGF method, which can improve the convergence of $v^2$ expansion comparing with the using of NRQCD factorization. 

Note, however, that both NRQCD factorization and SGF can also be used to directly describe low transverse momentum quarkonium production as well as quarkonium decay, where the QCD collinear factorization does not apply.

\section{Simplification}
\label{sec:sim}

\subsection{Collinear approximation}

The factorization formula in Eq.~\eqref{eq:fac4d} can be further simplified to make it more suitable in practice use.
In the rest frame of $\onia$, SGDs in Eq.~\eqref{eq:fac4d} have support only in the region $P^\mu=(M+O(\lambda^2/M),O(\lambda),$ $O(\lambda),O(\lambda))$, where $M^2=P^2$ and $\lambda\sim \text{a few hundreds MeV} <<M$ is the typical energy of emitted gluons. Thus we can expand $O(\lambda)$ and $O(\lambda^2/M)$ terms in hard parts. In this expansion, although space components of $P^{\mu}$ are $O(\lambda)$, the convergence should be much better than usual velocity expansion in NRQCD because $\vec{P}$ is integrated out symmetrically around the origin. The leading term in this expansion gives
\begin{align}\label{eq:fac1d}
(2\pi)^3 2 P_\onia^0 \frac{d\sigma_\onia}{d^3P_\onia}\approx \sum_{n} \int dz\, {\cal H}_{n}(P_\onia/z) F_{n\to \onia}(z),
\end{align}
where
\begin{align}
F_{n\to\onia}(z)=\int \frac{d^4P}{(2\pi)^4} \delta(z-\sqrt{P_\onia^2/P^2}) F_{n\to\onia}(P,P_\onia).
\end{align}
In this way, we significantly reduce the complication of input functions from four dimensional to one dimensional.  We refer to  Eq.~\eqref{eq:fac1d} as SGF-1d and  Eq.~\eqref{eq:fac4d} as SGF-4d.

Note that a similar formula that relates momentum of $\chi_{cJ}$ to its decaying particle $J/\psi$, $p_{J/\psi}\approx\frac{m_{J/\psi}}{m_{\chi_{cJ}}} p_{\chi_{cJ}}$, was first introduced by us in Ref.~\cite{Ma:2010vd}, and later also used in other publications \cite{Gong:2012ug,Bodwin:2015iua,Ma:2016exq}. The approximation was found to be very good, with only 8\% deviation from full results for both yield and polarization measurements \cite{Bodwin:2015iua}. We thus expect the approximation Eq.\eqref{eq:fac1d} to be also very good.

There are also other choices of collinear approximations. For any direction ${\hat{e}}$, we can have an approximation
\begin{align}
(2\pi)^3 2 P_\onia^0 \frac{d\sigma_\onia}{d^3P_\onia}\approx \sum_{n} \int dz\, {\cal H}_{n}(P_\onia/z) F^{\hat{e}}_{n\to \onia}(z),
\end{align}
with
\begin{align}
F^{\hat{e}}_{n\to\onia}(z)=\int \frac{d^4P}{(2\pi)^4} \delta(z-P_\onia\cdot {\hat{e}}/P\cdot {\hat{e}}) F_{n\to\onia}(P,P_\onia).
\end{align}
In the rest of the paper, we will only take Eq.~\eqref{eq:fac1d} as an example to show the effect of collinear approximation.

\subsection{Statical approximation}

As hard gluon emissions are excluded from $F_{n\to \onia}(z)$ and very soft gluon emissions are suppressed by phase space, $F_{n\to \onia}(z)$ should peak around $z=z_n\sim 1-O(\lambda/m_\onia)$. If the distribution of $F_{n\to \onia}(z)$ is narrow enough, we can approximate it by $\delta(z-z_n)\,\langle \widetilde{\mathcal O}_n^\onia \rangle$.
Then the SGF becomes
\begin{align}\label{eq:fac0d}
(2\pi)^3 2 P_\onia^0 \frac{d\sigma_\onia}{d^3P_\onia}\approx \sum_{n} {\cal H}_{n}(P_\onia/z_n) \langle \widetilde{\mathcal O}_n^\onia \rangle,
\end{align}
which we call SGF-0d. While this formula is as simple as NRQCD factorization, we will see later that the lowest order SGF-0d is a much better approximation of real physics than that of the lowest order NRQCD.

\subsection{Expansion of $m$}

In hard parts of all the above SGFs, there are at least two independent hard scales, $2m$ and $M$. Their difference is, however, much smaller than $2m$. We thus can further simplify hard parts by expanding $m$ around $M/2$, which defines a relativistic correction series in SGFs. We will see that there are no large corrections in this expansion.

\section{Example: $J/\psi$ hadroproduction}
\label{sec:jpsi}

\subsection{High $p_T$ quarkonium production}

To compare SGFs with NRQCD factorization, we apply them to the $J/\psi$ hadron production via gluon fragmentation. A charmonium $\onia$ production cross section via gluon fragmentation is given by
\begin{align}\label{eq:collinear}
d\sigma_{\onia}(p_T)= \int dx\, d\hat\sigma_g(p_T/x) D_{g\to \onia}(x),
\end{align}
where $d\hat\sigma_g(p_T/x)$ is the well-known hard part that produces a gluon with transverse momentum $p_T/x$, and $D_{g\to \onia}(x)$ is the FF of a gluon into an $\onia$ that will be determined in the following.

\subsection{Fragmentation functions calculated in SGF}

Using the SGF-4d Eq.~\eqref{eq:fac4d}, we obtain
\begin{align}\label{eq:FF4d}
D^{4d}_{g\to \onia}(x) = \sum_n \int \frac{dP^2}{2\pi} \frac{dy}{y} \hat{D}_n(y,P^2) \overline{F}_{n\to\onia}(\frac{x}{y},P^2),
\end{align}
where
\begin{align}
\overline{F}_{n\to\onia}(\frac{x}{y},P^2)=\int \frac{d^3P_\onia}{(2\pi)^32P_\onia^0} \delta(\frac{x}{y}-\frac{P_\onia^+}{P^+}) F_{n\to\onia}(P,P_\onia),
\end{align}
which only depends on $x/y$ and $P^2$ because of Lorentz symmetry combined with boost invariance along the ``+" direction.
Using the SGF-1d Eq.~\eqref{eq:fac1d}, we obtain
\begin{align}\label{eq:FF1d}
D^{1d}_{g\to \onia}(x) = \sum_n \int dz \hat{D}_n(\frac{x}{z},\frac{m_\onia^2}{z^2})\, z\, {F}_{n\to\onia}(z).
\end{align}

In Eqs.~\eqref{eq:FF4d} and \eqref{eq:FF1d}, $\hat{D}_n(y,P^2)$ can be calculated perturbatively. Up to $O(\alpha_s)$, only the $n=\COcSa$ channel has a nonvanishing contribution \cite{machao},
\begin{align}\label{eq:hard}
\hat{D}_{\COcSa}^{(1)}(x,M^2)=\frac{\pi \alpha_s}{24 m^3}\frac{\left(\frac{1+2\Delta}{3}\right)^2}{\Delta^5} \delta(1-x),
\end{align}
with $\Delta=\frac{M}{2m}$. Equation \eqref{eq:hard} can also be used in NRQCD factorization, where one expresses $\Delta=\sqrt{1+\langle v\rangle^2}$ and then expands $\langle v\rangle^2$, which gives a normalized series $1-\frac{11}{6}\langle v\rangle^2+\frac{191}{72}\langle v\rangle^4+\cdots$. The first two terms in this expansion have been calculated before~\cite{Braaten:1994kd,Bodwin:2003wh}.

\subsection{Model assumptions}

For the nonperturbative function $F_{n\to\onia}(P,P_\onia)$, we simply employ an existing model in Ref.~\cite{Beneke:1999gq},
\begin{align}
F_{\COcSa\to\onia}(P,P_\onia) = a\, k^2\, \text{exp}(-\frac{k_0^2+k^2}{\Lambda^2}),
\end{align}
where $k=P-P_\onia$ is a timelike momentum with positive energy $k_0$. $\Lambda$ is an energy cutoff for emitted soft gluons, for which we will choose $500$MeV \footnote{By assuming the cutoff at the order of $mv^2$, this choice corresponds to $v^2\approx0.3$. As we will see, $D^{1d}_{g\to \onia}(x)$ calculated in this model is picked around $x\approx1-v^2/2$.}. As we are only interested in the cross section ratio in the following, the value of $a$ is irrelevant. We further set $m=1.55$GeV in studying $J/\psi$ production. We note that, our conclusions are in fact independent of these choices.

\begin{figure}[htb!]
 \begin{center}
 \includegraphics[width=0.35\textwidth]{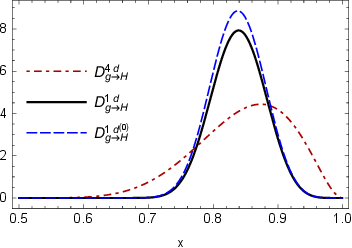}
  \caption{ Fragmentation functions calculated by SGFs. The overall normalization is arbitrary. \label{fig:FFs}}
 \end{center}
\end{figure}

\subsection{Numerical results}

With these inputs in hand, FFs calculated by using different methods are shown in Fig.~\ref{fig:FFs}. In this figure, $D^{4d}_{g\to \onia}(x)$ and $D^{1d}_{g\to \onia}(x)$ are calculated by using the exact value of $m$, while $D^{1d^{(0)}}_{g\to \onia}(x)$ is similar to $D^{1d}_{g\to \onia}(x)$ but calculated by setting $m=M/2$ in Eq.~\eqref{eq:hard}. We find that FFs calculated by using SGF-4d and SGF-1d have different shapes, but they have the same accumulated values,
\begin{align}
\int_0^1 dx\, D^{4d}_{g\to \onia}(x) = \int_0^1 dx\, D^{1d}_{g\to \onia}(x).
\end{align}
 This is because SGF-4d and SGF-1d are equivalent for the integrated FF. The FF obtained by expanding $m$ to the lowest order, $D^{1d^{(0)}}_{g\to \onia}(x)$, has only a small difference from the complete result $D^{1d}_{g\to \onia}(x)$, which implies that relativistic correction due to the expansion of $m$ is very small.

We also find that $F_{\COcSa\to\onia}(z)$ has almost the same shape as that of $D^{1d}_{g\to \onia}(x)$, which is $\delta$-function-like. For cross section, we will find that SGF-0d with $z_0=0.86$ can well reproduce SGF-4d.

\begin{figure}[htb!]
 \begin{center}
 \includegraphics[width=0.4\textwidth]{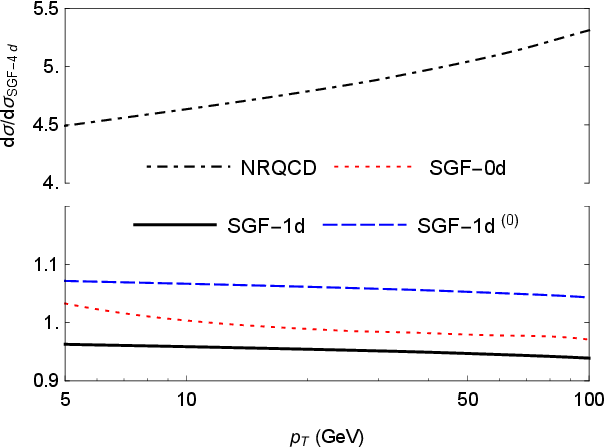}
  \caption{Ratio of $J/\psi$ differential cross section at LHC calculated from different approaches over that calculated from SGF-4d. See the text for details.
   \label{fig:xsec}}
 \end{center}
\end{figure}

Using the above FFs, we can calculate the $J/\psi$ cross section based on Eq.~\eqref{eq:collinear}. Let us assume that the cross section calculated by SGF-4d is ``exact'', and examine how good are the SGF-1d, SGF-0d, and NRQCD expansions.
We show ratios of $J/\psi$ differential cross section calculated by different methods over that calculated by SGF-4d in Fig.~\ref{fig:xsec}. We find that the SGF-1d result is close to the exact value, with largest error less than 6\%. This implies that, as expected, the SGF-1d expansion should be very good. By expanding $m$ to the lowest order, SGF-1d$^{(0)}$ also provides a good approximation, with deviation smaller than 10\%. The SGF-0d can well reproduce SGF-4d if we choose $z_0=0.86$, as shown in Fig.~\ref{fig:xsec}. So, we expect that the convergence of velocity expansion in SGF is good in general.

On the contrary, the lowest order NRQCD result is larger than the exact value by more than a factor of 4, where NRQCD LDMEs are determined by the approximation Eq.~\eqref{eq:ldmes}. There are two main sources for the large deviation. One comes from the hard part of Eq.~\eqref{eq:collinear}, which has an approximate scaling behavior $d\hat\sigma_g(p_T/x)\sim (p_T/x)^{-4}$ with average value of $x$ being around 0.86. Yet, by ignoring soft gluon emission, NRQCD approximates $x\approx1$, which enhances the total result by almost a factor of 2. This effect has also been pointed out in Sec.\ref{sec:badconv}. The other one comes from Eq.~\eqref{eq:hard}. By ignoring soft gluon emission and then expanding $M$ around $2m$, NRQCD approximates $\Delta\approx\frac{m_{J/\psi}/0.86}{2m}\approx1.16$ by 1, which enhances the total result by another factor of 2. Roughly speaking, the lowest order NRQCD approximates $0.86^9\approx1/4\sim (1-v^2/2)^9$ by 1 in this problem, which is hard to be recovered by traditional relativistic correction in NRQCD~\cite{Xu:2012am}.

For bottomonia production, because $v^2$ is smaller, NRQCD approximation can be a little bit better than that for charmonia. According to the above discussion, we expect the lowest order NRQCD approximation overestimates bottomonia production rate by a factor of $(1-v^2/2)^{-9}\approx1.6$.

\section{Summary and outlook}
\label{sec:sum}

In this paper, we propose a soft gluon factorization (SGF) approach to describe quarkonium production and decay, which keeps the momentum difference between intermediate $\qq$ pair and physical quarkonium. From the point view of NRQCD, the SGF effectively resums a subset of Lorentz-invariant relativistic correction terms in NRQCD factorization. In this sense, the SGF is a generalization of the NRQCD factorization.

By construction, the SGF has a good convergence in velocity expansion, so  the lowest order approximation may already capture most physics; whereas for NRQCD factorization, we find that the lowest order result in velocity expansion can deviate from the full result by more than a factor of 4, mainly due to ignoring the momentum taken away by soft gluon emission.
With so large deviation, it is not surprising that the NRQCD calculation faces many difficulties.

Hopefully, these difficulties may be resolved or relieved in the SGF framework with well controlled relativistic corrections. Specifically, the universality problem may be due to the fact that, for quarkonium production in different processes, e.g., $e^+e^-$ collision or $pp$ collision, lowest order NRQCD calculations suffer from large but different  relativistic corrections. While in SGF, we do not expect large relativistic corrections.
Moreover, considering the large relativistic effect on yield, we may also expect a significant effect on polarization, like color-magnetic dipole transition effect, which may alter the transverse polarization of the $\COcSa$ channel. Further examinations should be performed on the above expectations.

\begin{acknowledgments}

We thank G. Bodwin, A.P. Chen, C. Meng, J.W. Qiu, K. Wang, F. Yuan, H. Zhang and Y.J. Zhang for many useful communications and discussions. In particular, we thank J.Z. Li for very helpful discussions in the earlier stage of this work.
The work is supported in part by the National Natural Science Foundation
of China (Grants No. 11975029, No. 11475005, No. 11875071 and No. 11075002) and the National Key Basic Research Program of China (No. 2015CB856700).

\appendix

\section{Definition of angular momentum operators}
\label{sec:defangular}

Angular momentum operators in Eq.~\eqref{eq:Kn} are defined as
\begin{align}
\Gamma_{n}=\sum_{L_z,S_z} \langle L, L_z; S,S_z| J,J_z \rangle \Gamma^o_{LL_z}\Gamma^s_{SS_z} \,.
\end{align}
As total spin has only two choices, $S=0$ or $1$, the corresponding spin operators are
\begin{subequations}\label{eq:spinProj}
	\begin{align}
		{\Gamma}^s_{00} &= \gamma_5,\\
		{\Gamma}^s_{1S_z} &= {\epsilon}_{S_z}^\mu \gamma_\mu,
	\end{align}
\end{subequations}
where ${\epsilon}_{S_z}^\mu$ are polarization vectors. Orbital operators are defined as
\begin{align}\label{eq:orbital}
{\Gamma}^o_{L L_z} &=  {\epsilon}_{L_z}^{\mu_1\cdots\mu_L} (-\frac{i}{2})^L\overleftrightarrow{D}_{\hspace{-1mm}\mu_1}\cdots\overleftrightarrow{D}_{\hspace{-1mm}\mu_L},
\end{align}
where ${D}_\mu$ is the gauge covariant derivative with $\bar\Psi\overleftrightarrow{D}_{\hspace{-1mm}\mu}\Psi=\bar\Psi\left({D}_\mu\Psi\right)-\left(D_\mu\bar\Psi\right)\Psi$ and ${\epsilon}_{L_z}^{\mu_1\cdots\mu_L} $ are $L$-rank polarization tensors.
The above polarization tensors ${\epsilon}_{S_z}^\mu$ and ${\epsilon}_{L_z}^{\mu_1\cdots\mu_L}$ have only spatial components in the rest frame of $P_{\onia\mu}$, which is equivalent to the following relations:
\begin{subequations}
	\begin{align}
        P_\onia \cdot {\epsilon}_{S_z}&=0,\\
        P_{\onia\mu_i} \cdot {\epsilon}_{L_z}^{\mu_1\cdots\mu_L}&=0~~~~\text{for}~~~~i=1,\cdots,L.
	\end{align}
\end{subequations}
Furthermore, polarization tensors are chosen to be orthonormal, e.g., ${\epsilon}_{S_z}\cdot {\epsilon}_{S_z'}^*=-\delta_{S_zS_z'}$. The combination of the above two properties results in
\begin{align}
	\sum_{S_z} \epsilon_{S_z}^\mu\epsilon_{S_z}^{*\nu}=-g^{\mu\nu}+\frac{P_\onia^\mu P_\onia^\nu}{M_\onia^2}\,.
\end{align}

Although orbital operators in Eq.~\eqref{eq:orbital} are similar to that in NRQCD, their meanings are actually very different. In fact, in addition to these operators, there are an infinite number of relativistic-correction operators in NRQCD, like $(-\frac{i}{2})^2 \overleftrightarrow{D}^2$, which do not show up in SGF. The reason is that, instead of relativistic expansion, the SGF is performing a partial-wave expansion. This can be seen more clearly in momentum space. Then in the rest frame of $P_\onia$, Eq.~\eqref{eq:orbital} becomes
\begin{align}\label{eq:momEps}
	{\epsilon}_{L_z}^{\mu_1\cdots\mu_L} q_{\mu_1}\cdots q_{\mu_L} &= |{\bm q}|^L \sqrt{\frac{4\pi (L!)}{(2L+1)!!}}Y_{L}^{L_z}(\theta, \phi),
\end{align}
where $q_\mu$ is half of the relative momentum between intermediate $\qq$, and $\theta$ and $\phi$ are polar angle and azimuthal angle of $\bm q$, respectively. Equation~\eqref{eq:momEps} can be thought of as the definition of ${\Gamma}^o_{L L_z}$ in momentum space. As the partial-wave expansion already forms a complete set of operators, we do not need further relativistic-correction operators. Relativistic corrections in SGF are encoded in perturbative calculable hard parts.  Our definition of SGDs is to pick up the minimal number of $|{\bm q}|$ in each partial wave.

\begin{figure}[tbh!]
	\begin{center}
		\includegraphics[width=0.25\textwidth]{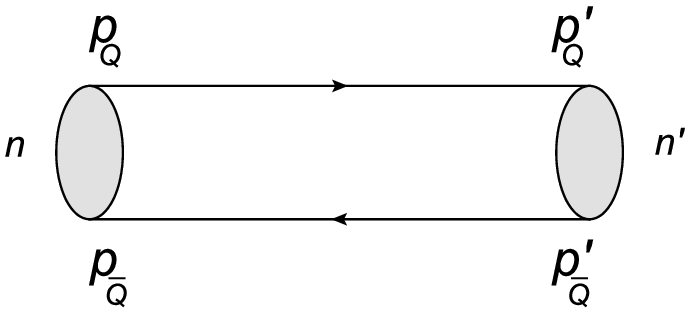}
		\caption{ Diagram of $F_{n\to n'}^{(0)}(P,P')$. \label{fig:F0}}
	\end{center}
\end{figure}

\section{Leading order expansion of SGDs}
\label{sec:checkLO}

The Feynman diagram which represents the leading order expansion $F_{n\to n'}^{(0)}(P,P')$ is shown in Fig.~\ref{fig:F0}.
It is clear that, if the color of $n$ is different from that of $n'$, then the amplitude vanishes. If both $n$ and $n'$ are color singlet, we get ${\text{Tr}}_c[\frac{1}{\sqrt{N_c}}\frac{1}{\sqrt{N_c}}]=1$. If both $n$ and $n'$ are color octet, for amplitude we have ${\text{Tr}}_c[\sqrt{2}t^a \sqrt\frac{2}{N_c^2-1} t^{a'}]=\frac{\delta_{aa'}}{\sqrt{N_c^2-1}}$, which results in $\frac{\delta_{aa'}}{\sqrt{N_c^2-1}}\frac{\delta_{aa'}}{\sqrt{N_c^2-1}}=1$ after summing over color states. Thus for color factor we always have $\delta_{cc'}$.

The angular momentum part of the amplitude gives
\begin{widetext}
\begin{align}
\begin{split}
A_{n\to n'}^{(0)}&(P,P')=\,\left[\sum_{L_z,S_z} \langle L, L_z; S,S_z|J,J_z \rangle \int \frac{d^4q}{(2\pi)^4}|{\bm q}|^L \sqrt{\frac{4\pi (L!)}{(2L+1)!!}} Y_{L}^{L_z}(\theta, \phi) \right] \,\\
&\times \left[\sum_{L_z^\prime,S_z^\prime} \langle L^\prime, L_z^\prime; S^\prime,S_z^\prime | J^\prime,J_z^\prime \rangle \int d^2 \Omega^\prime |{\bm q}'|^{-L'} \sqrt{\frac{(2L'+1)!!}{4\pi (L'!)}}  Y_{L^\prime}^{*L_z^\prime}(\theta', \phi')\right] (2\pi)^4 \delta^4(p_Q-p_{Q}') \\
&\times   \frac{2\sqrt{M'}}{\sqrt{M'}(M'+2m)^2} \text{Tr}\left[  \frac{M'+\slashed{P}'}{2M'} \Gamma^s_{SS_z} \frac{M'-\slashed{P}'}{2M'}  (\slashed{p}_{\overline Q}'-m) \frac{M'-\slashed{P}'}{2M'} \widetilde{\Gamma}^s_{S'S_z'} \frac{M'+\slashed{P}'}{2M'} (\slashed{p}_{ Q}'+m) \right],\\
\end{split}
\end{align}
where $(2\pi)^4 \delta^4(p_Q-p_{Q}')$ presents because the diagram is disconnected and the terms inside of ``Tr" project both initial $\qq$ and final $\qq$ to specific spin states. By setting $P=P'$ and using the delta function to integrate out $q$, we get
\begin{align}\label{eq:A02}
\begin{split}
A_{n\to n'}^{(0)}(&P,P')=\sum_{L_z,S_z,L_z^\prime,S_z^\prime} \langle L, L_z; S,S_z | J,J_z\rangle \langle L^\prime, L_z^\prime; S^\prime,S_z^\prime|J^\prime,J_z^\prime \rangle \int d^2 \Omega^\prime Y_{L}^{L_z}(\theta^\prime, \phi^\prime) Y_{L^\prime}^{*L_z^\prime}(\theta^\prime, \phi^\prime)  \,|{\bm q}|^{L-L'}\\
&\times   \sqrt{\frac{(2L'+1)!! (L!)}{(2L+1)!!(L'!)}} \frac{2}{(M+2m)^2} \text{Tr}\left[  \frac{M+\slashed{P}}{2M} \Gamma^{s}_{SS_z} \frac{M-\slashed{P}}{2M}  (\slashed{p}_{\overline Q}-m) \frac{M-\slashed{P}}{2M} \widetilde{\Gamma}^s_{S'S_z'} \frac{M+\slashed{P}}{2M} (\slashed{p}_{ Q}+m) \right]\\
=&\delta_{LL'} \sum_{L_z,S_z,S_z^\prime} \langle L, L_z; S,S_z | J,J_z \rangle \langle L, L_z; S^\prime,S_z^\prime| J^\prime,J_z^\prime \rangle  \,\\
&\times  \frac{2}{(M+2m)^2} \text{Tr}\left[  \frac{M+\slashed{P}}{2M} \Gamma^{s}_{SS_z} \frac{M-\slashed{P}}{2M}  (\slashed{p}_{\overline Q}-m) \frac{M-\slashed{P}}{2M} \widetilde{\Gamma}^s_{S'S_z'} \frac{M+\slashed{P}}{2M} (\slashed{p}_{ Q}+m) \right].
\end{split}
\end{align}
where we have used $\int d^2 \Omega^\prime Y_{L}^{L_z}(\theta^\prime, \phi^\prime) Y_{L^\prime}^{*L_z^\prime}(\theta^\prime, \phi^\prime)=\delta_{LL'}\delta_{L_zL_z'}$ in the last step.
Due to
\begin{subequations}
	\begin{align}
\frac{M-\slashed{P}}{2M}  (\slashed{p}_{\overline Q}-m) \frac{M-\slashed{P}}{2M}=&-(m+\frac{M}{2})\frac{M-\slashed{P}}{2M},\\
\frac{M+\slashed{P}}{2M} (\slashed{p}_{ Q}+m)\frac{M+\slashed{P}}{2M} =&(m+\frac{M}{2})\frac{M+\slashed{P}}{2M},
\end{align}
\end{subequations}
the last line of Eq.~\eqref{eq:A02} gives
\begin{align}
\begin{split}
 -\frac{1}{2} \text{Tr}\left[  \frac{M+\slashed{P}}{2M} \Gamma^s_{SS_z} \frac{M-\slashed{P}}{2M} \widetilde{\Gamma}^s_{S'S_z'} \right]=-\frac{1}{2} \text{Tr}\left[  \frac{M+\slashed{P}}{2M} \Gamma^s_{SS_z}  \widetilde{\Gamma}^s_{S'S_z'} \right]=\delta_{SS'}\delta_{S_zS_z'},
\end{split}
\end{align}
where we have used the fact that $P_\onia=P$ at this order and thus $P\cdot \epsilon_{S_z}=0$. Therefore,
\begin{align}
\begin{split}
A_{n\to n'}^{(0)}(P,P')=&\delta_{LL'}\delta_{SS'}\sum_{L_z,S_z} \langle L, L_z; S,S_z|J,J_z \rangle \langle L, L_z; S,S_z|J^\prime,J_z^\prime  \rangle=\delta_{LL'}\delta_{SS'}\delta_{JJ'}\delta_{J_zJ_z'}.
\end{split}
\end{align}
Eventually, we have
\begin{align}
\begin{split}
F_{n\to n'}^{(0)}(P,P')=(2\pi)^4\delta^4(P-P')\delta_{cc'}|A_{n\to n'}^{(0)}(P,P')|^2=(2\pi)^4\delta^4(P-P')\delta_{nn'}.
\end{split}
\end{align}

\end{widetext}
\end{acknowledgments}

\providecommand{\href}[2]{#2}\begingroup\raggedright\endgroup


\begin{thebibliography}{10}

\bibitem{Ellis:1976fj}
S.~Ellis, M.~B. Einhorn, and C.~Quigg, {\it {Comment on Hadronic Production of
  Psions}},
\href{http://dx.doi.org/10.1103/PhysRevLett.36.1263}{{\em Phys.Rev.Lett.}
  {\bfseries 36} (1976) 1263}
  [\href{http://inspirehep.net/search?p=find+Ellis:1976fj}{{\ttfamily
  InSPIRE}}].

\bibitem{Carlson:1976cd}
C.~Carlson and R.~Suaya, {\it {Hadronic Production of $\psi/J$ Mesons}},
\href{http://dx.doi.org/10.1103/PhysRevD.14.3115}{{\em Phys.Rev.} {\bfseries
  D14} (1976) 3115}
  [\href{http://inspirehep.net/search?p=find+Carlson:1976cd}{{\ttfamily
  InSPIRE}}].

\bibitem{Chang:1979nn}
C.-H. Chang, {\it {Hadronic Production of $J/\psi$ Associated With a Gluon}},
\href{http://dx.doi.org/10.1016/0550-3213(80)90175-3}{{\em Nucl.Phys.}
  {\bfseries B172} (1980) 425--434}
  [\href{http://inspirehep.net/search?p=find+Chang:1979nn}{{\ttfamily
  InSPIRE}}].

\bibitem{Bodwin:1994jh}
G.~T. Bodwin, E.~Braaten, and G.~P. Lepage, {\it {Rigorous QCD analysis of
  inclusive annihilation and production of heavy quarkonium}},
  \href{http://dx.doi.org/10.1103/PhysRevD.55.5853,
  10.1103/PhysRevD.51.1125}{{\em Phys. Rev.} {\bfseries D51} (1995) 1125--1171}
  [\href{http://arxiv.org/abs/hep-ph/9407339}{{\ttfamily hep-ph/9407339}}]
  [\href{http://inspirehep.net/search?p=find+Bodwin:1994jh}{{\ttfamily
  InSPIRE}}].
[Erratum: Phys. Rev.D55,5853(1997)].

\bibitem{Fritzsch:1977ay}
H.~Fritzsch, {\it {Producing Heavy Quark Flavors in Hadronic Collisions: A Test
  of Quantum Chromodynamics}},
\href{http://dx.doi.org/10.1016/0370-2693(77)90108-3}{{\em Phys.Lett.}
  {\bfseries B67} (1977) 217}
  [\href{http://inspirehep.net/search?p=find+Fritzsch:1977ay}{{\ttfamily
  InSPIRE}}].

\bibitem{Halzen:1977rs}
F.~Halzen, {\it {Cvc for Gluons and Hadroproduction of Quark Flavors}},
\href{http://dx.doi.org/10.1016/0370-2693(77)90144-7}{{\em Phys.Lett.}
  {\bfseries B69} (1977) 105}
  [\href{http://inspirehep.net/search?p=find+Halzen:1977rs}{{\ttfamily
  InSPIRE}}].

\bibitem{Ma:2016exq}
Y.-Q. Ma and R.~Vogt, {\it {Quarkonium Production in an Improved Color
  Evaporation Model}},
\href{http://dx.doi.org/10.1103/PhysRevD.94.114029}{{\em Phys. Rev.} {\bfseries
  D94} (2016) 114029} [\href{http://arxiv.org/abs/1609.06042}{{\ttfamily
  arXiv:1609.06042}}]
  [\href{http://inspirehep.net/search?p=find+Ma:2016exq}{{\ttfamily InSPIRE}}].

\bibitem{Kang:2011mg}
Z.-B. Kang, J.-W. Qiu, and G.~Sterman, {\it {Heavy quarkonium production and
  polarization}},
\href{http://dx.doi.org/10.1103/PhysRevLett.108.102002}{{\em Phys.Rev.Lett.}
  {\bfseries 108} (2012) 102002}
  [\href{http://arxiv.org/abs/1109.1520}{{\ttfamily arXiv:1109.1520}}]
  [\href{http://inspirehep.net/search?p=find+Kang:2011mg}{{\ttfamily
  InSPIRE}}].

\bibitem{Fleming:2012wy}
S.~Fleming, A.~K. Leibovich, T.~Mehen, and I.~Z. Rothstein, {\it {The
  Systematics of Quarkonium Production at the LHC and Double Parton
  Fragmentation}},
\href{http://dx.doi.org/10.1103/PhysRevD.86.094012}{{\em Phys.Rev.} {\bfseries
  D86} (2012) 094012} [\href{http://arxiv.org/abs/1207.2578}{{\ttfamily
  arXiv:1207.2578}}]
  [\href{http://inspirehep.net/search?p=find+Fleming:2012wy}{{\ttfamily
  InSPIRE}}].

\bibitem{Kang:2014tta}
Z.-B. Kang, Y.-Q. Ma, J.-W. Qiu, and G.~Sterman, {\it {Heavy quarkonium
  production at collider energies: Factorization and Evolution}},
\href{http://dx.doi.org/10.1103/PhysRevD.90.034006}{{\em Phys.Rev.} {\bfseries
  D90} (2014) 034006} [\href{http://arxiv.org/abs/1401.0923}{{\ttfamily
  arXiv:1401.0923}}]
  [\href{http://inspirehep.net/search?p=find+Kang:2014tta}{{\ttfamily
  InSPIRE}}].

\bibitem{Bodwin:1992qr}
G.~T. Bodwin, E.~Braaten, T.~C. Yuan, and G.~P. Lepage, {\it {P wave charmonium
  production in B meson decays}},
\href{http://dx.doi.org/10.1103/PhysRevD.46.R3703}{{\em Phys.Rev.} {\bfseries
  D46} (1992) 3703--3707}
  [\href{http://arxiv.org/abs/hep-ph/9208254}{{\ttfamily hep-ph/9208254}}]
  [\href{http://inspirehep.net/search?p=find+Bodwin:1992qr}{{\ttfamily
  InSPIRE}}].

\bibitem{Braaten:1994vv}
E.~Braaten and S.~Fleming, {\it {Color octet fragmentation and the
  $\psi^\prime$ surplus at the Tevatron}},
\href{http://dx.doi.org/10.1103/PhysRevLett.74.3327}{{\em Phys.Rev.Lett.}
  {\bfseries 74} (1995) 3327--3330}
  [\href{http://arxiv.org/abs/hep-ph/9411365}{{\ttfamily hep-ph/9411365}}]
  [\href{http://inspirehep.net/search?p=find+Braaten:1994vv}{{\ttfamily
  InSPIRE}}].

\bibitem{Boyd:1998km}
C.~G. Boyd, A.~K. Leibovich, and I.~Rothstein, {\it {$J\psi$ production at LEP:
  Revisited and resummed}},
\href{http://dx.doi.org/10.1103/PhysRevD.59.054016}{{\em Phys.Rev.} {\bfseries
  D59} (1999) 054016} [\href{http://arxiv.org/abs/hep-ph/9810364}{{\ttfamily
  hep-ph/9810364}}]
  [\href{http://inspirehep.net/search?p=find+Boyd:1998km}{{\ttfamily
  InSPIRE}}].

\bibitem{Kramer:2001hh}
M.~Kramer, {\it {Quarkonium production at high-energy colliders}},
\href{http://dx.doi.org/10.1016/S0146-6410(01)00154-5}{{\em
  Prog.Part.Nucl.Phys.} {\bfseries 47} (2001) 141--201}
  [\href{http://arxiv.org/abs/hep-ph/0106120}{{\ttfamily hep-ph/0106120}}]
  [\href{http://inspirehep.net/search?p=find+Kramer:2001hh}{{\ttfamily
  InSPIRE}}].

\bibitem{Klasen:2001cu}
M.~Klasen, B.~A. Kniehl, L.~Mihaila, and M.~Steinhauser, {\it {Evidence for
  color octet mechanism from CERN LEP-2 $\gamma \gamma \to J/\psi$ + $X$
  data}},
\href{http://dx.doi.org/10.1103/PhysRevLett.89.032001}{{\em Phys.Rev.Lett.}
  {\bfseries 89} (2002) 032001}
  [\href{http://arxiv.org/abs/hep-ph/0112259}{{\ttfamily hep-ph/0112259}}]
  [\href{http://inspirehep.net/search?p=find+Klasen:2001cu}{{\ttfamily
  InSPIRE}}].

\bibitem{Zhang:2006ay}
Y.-J. Zhang and K.-T. Chao, {\it {Double charm production $e^+ e^- \to J/\psi +
  c \bar{c}$ at B factories with next-to-leading order QCD correction}},
\href{http://dx.doi.org/10.1103/PhysRevLett.98.092003}{{\em Phys.Rev.Lett.}
  {\bfseries 98} (2007) 092003}
  [\href{http://arxiv.org/abs/hep-ph/0611086}{{\ttfamily hep-ph/0611086}}]
  [\href{http://inspirehep.net/search?p=find+Zhang:2006ay}{{\ttfamily
  InSPIRE}}].

\bibitem{Lansberg:2008gk}
J.~Lansberg, {\it {On the mechanisms of heavy-quarkonium hadroproduction}},
\href{http://dx.doi.org/10.1140/epjc/s10052-008-0826-9}{{\em Eur.Phys.J.}
  {\bfseries C61} (2009) 693--703}
  [\href{http://arxiv.org/abs/0811.4005}{{\ttfamily arXiv:0811.4005}}]
  [\href{http://inspirehep.net/search?p=find+Lansberg:2008gk}{{\ttfamily
  InSPIRE}}].

\bibitem{Ma:2008gq}
Y.-Q. Ma, Y.-J. Zhang, and K.-T. Chao, {\it {QCD correction to $e^+ e^- \to
  J/\psi + g + g$ at B Factories}},
\href{http://dx.doi.org/10.1103/PhysRevLett.102.162002}{{\em Phys.Rev.Lett.}
  {\bfseries 102} (2009) 162002}
  [\href{http://arxiv.org/abs/0812.5106}{{\ttfamily arXiv:0812.5106}}]
  [\href{http://inspirehep.net/search?p=find+Ma:2008gq}{{\ttfamily InSPIRE}}].

\bibitem{Gong:2009kp}
B.~Gong and J.-X. Wang, {\it {Next-to-Leading-Order QCD Corrections to $e^+ e^-
  \to J/\psi +g+g$ at the B Factories}},
\href{http://dx.doi.org/10.1103/PhysRevLett.102.162003}{{\em Phys.Rev.Lett.}
  {\bfseries 102} (2009) 162003}
  [\href{http://arxiv.org/abs/0901.0117}{{\ttfamily arXiv:0901.0117}}]
  [\href{http://inspirehep.net/search?p=find+Gong:2009kp}{{\ttfamily
  InSPIRE}}].

\bibitem{Butenschoen:2009zy}
M.~Butenschoen and B.~A. Kniehl, {\it {Complete next-to-leading-order
  corrections to $J/\psi$ photoproduction in nonrelativistic quantum
  chromodynamics}},
\href{http://dx.doi.org/10.1103/PhysRevLett.104.072001}{{\em Phys.Rev.Lett.}
  {\bfseries 104} (2010) 072001}
  [\href{http://arxiv.org/abs/0909.2798}{{\ttfamily arXiv:0909.2798}}]
  [\href{http://inspirehep.net/search?p=find+Butenschoen:2009zy}{{\ttfamily
  InSPIRE}}].

\bibitem{Ma:2010vd}
Y.-Q. Ma, K.~Wang, and K.-T. Chao, {\it {QCD radiative corrections to
  $\chi_{cJ}$ production at hadron colliders}},
\href{http://dx.doi.org/10.1103/PhysRevD.83.111503}{{\em Phys.Rev.} {\bfseries
  D83} (2011) 111503} [\href{http://arxiv.org/abs/1002.3987}{{\ttfamily
  arXiv:1002.3987}}]
  [\href{http://inspirehep.net/search?p=find+Ma:2010vd}{{\ttfamily InSPIRE}}].

\bibitem{Ma:2010yw}
Y.-Q. Ma, K.~Wang, and K.-T. Chao, {\it {$J/\psi (\psi^\prime)$ production at
  the Tevatron and LHC at ${\cal O}(\alpha_s^4v^4)$ in nonrelativistic QCD}},
\href{http://dx.doi.org/10.1103/PhysRevLett.106.042002}{{\em Phys.Rev.Lett.}
  {\bfseries 106} (2011) 042002}
  [\href{http://arxiv.org/abs/1009.3655}{{\ttfamily arXiv:1009.3655}}]
  [\href{http://inspirehep.net/search?p=find+Ma:2010yw}{{\ttfamily InSPIRE}}].

\bibitem{Butenschoen:2010rq}
M.~Butenschoen and B.~A. Kniehl, {\it {Reconciling $J/\psi$ production at HERA,
  RHIC, Tevatron, and LHC with NRQCD factorization at next-to-leading order}},
\href{http://dx.doi.org/10.1103/PhysRevLett.106.022003}{{\em Phys.Rev.Lett.}
  {\bfseries 106} (2011) 022003}
  [\href{http://arxiv.org/abs/1009.5662}{{\ttfamily arXiv:1009.5662}}]
  [\href{http://inspirehep.net/search?p=find+Butenschoen:2010rq}{{\ttfamily
  InSPIRE}}].

\bibitem{Ma:2014mri}
Y.-Q. Ma and R.~Venugopalan, {\it {Comprehensive Description of $J/\psi$
  Production in Proton-Proton Collisions at Collider Energies}},
\href{http://dx.doi.org/10.1103/PhysRevLett.113.192301}{{\em Phys.Rev.Lett.}
  {\bfseries 113} (2014) 192301}
  [\href{http://arxiv.org/abs/1408.4075}{{\ttfamily arXiv:1408.4075}}]
  [\href{http://inspirehep.net/search?p=find+Ma:2014mri}{{\ttfamily InSPIRE}}].

\bibitem{Nayak:2005rt}
G.~C. Nayak, J.-W. Qiu, and G.~Sterman, {\it {Fragmentation, NRQCD and NNLO
  factorization analysis in heavy quarkonium production}},
\href{http://dx.doi.org/10.1103/PhysRevD.72.114012}{{\em Phys.Rev.} {\bfseries
  D72} (2005) 114012} [\href{http://arxiv.org/abs/hep-ph/0509021}{{\ttfamily
  hep-ph/0509021}}]
  [\href{http://inspirehep.net/search?p=find+Nayak:2005rt}{{\ttfamily
  InSPIRE}}].

\bibitem{Nayak:2005rw}
G.~C. Nayak, J.-W. Qiu, and G.~Sterman, {\it {Fragmentation, factorization and
  infrared poles in heavy quarkonium production}},
\href{http://dx.doi.org/10.1016/j.physletb.2005.03.031}{{\em Phys.Lett.}
  {\bfseries B613} (2005) 45--51}
  [\href{http://arxiv.org/abs/hep-ph/0501235}{{\ttfamily hep-ph/0501235}}]
  [\href{http://inspirehep.net/search?p=find+Nayak:2005rw}{{\ttfamily
  InSPIRE}}].

\bibitem{Nayak:2006fm}
G.~C. Nayak, J.-W. Qiu, and G.~Sterman, {\it {NRQCD Factorization and
  Velocity-dependence of NNLO Poles in Heavy Quarkonium Production}},
\href{http://dx.doi.org/10.1103/PhysRevD.74.074007}{{\em Phys.Rev.} {\bfseries
  D74} (2006) 074007} [\href{http://arxiv.org/abs/hep-ph/0608066}{{\ttfamily
  hep-ph/0608066}}]
  [\href{http://inspirehep.net/search?p=find+Nayak:2006fm}{{\ttfamily
  InSPIRE}}].

\bibitem{Cho:1994ih}
P.~L. Cho and M.~B. Wise, {\it {Spin symmetry predictions for heavy quarkonia
  alignment}},
\href{http://dx.doi.org/10.1016/0370-2693(94)01658-Y}{{\em Phys. Lett.}
  {\bfseries B346} (1995) 129--136}
  [\href{http://arxiv.org/abs/hep-ph/9411303}{{\ttfamily hep-ph/9411303}}]
  [\href{http://inspirehep.net/search?p=find+Cho:1994ih}{{\ttfamily InSPIRE}}].

\bibitem{Beneke:1995yb}
M.~Beneke and I.~Z. Rothstein, {\it {Psi-prime polarization as a test of color
  octet quarkonium production}},
  \href{http://dx.doi.org/10.1016/S0370-2693(96)80022-0,
  10.1016/0370-2693(96)00030-5}{{\em Phys. Lett.} {\bfseries B372} (1996)
  157--164} [\href{http://arxiv.org/abs/hep-ph/9509375}{{\ttfamily
  hep-ph/9509375}}]
  [\href{http://inspirehep.net/search?p=find+Beneke:1995yb}{{\ttfamily
  InSPIRE}}].
[Erratum: Phys. Lett.B389,769(1996)].

\bibitem{Braaten:1999qk}
E.~Braaten, B.~A. Kniehl, and J.~Lee, {\it {Polarization of prompt $J/\psi$ at
  the Tevatron}},
\href{http://dx.doi.org/10.1103/PhysRevD.62.094005}{{\em Phys.Rev.} {\bfseries
  D62} (2000) 094005} [\href{http://arxiv.org/abs/hep-ph/9911436}{{\ttfamily
  hep-ph/9911436}}]
  [\href{http://inspirehep.net/search?p=find+Braaten:1999qk}{{\ttfamily
  InSPIRE}}].

\bibitem{Abulencia:2007us}
{\bfseries CDF Collaboration} , A.~Abulencia {\em et al.}, {\it {Polarization
  of $J/\psi$ and $\psi(2S)$ mesons produced in $p \bar{p}$ collisions at
  $\sqrt{s}$ = 1.96-TeV}},
\href{http://dx.doi.org/10.1103/PhysRevLett.99.132001}{{\em Phys.Rev.Lett.}
  {\bfseries 99} (2007) 132001}
  [\href{http://arxiv.org/abs/0704.0638}{{\ttfamily arXiv:0704.0638}}]
  [\href{http://inspirehep.net/search?p=find+Abulencia:2007us}{{\ttfamily
  InSPIRE}}].

\bibitem{Abelev:2011md}
{\bfseries ALICE Collaboration} , B.~Abelev {\em et al.}, {\it {$J/\psi$
  polarization in $pp$ collisions at $\sqrt{s}=7$ TeV}},
\href{http://dx.doi.org/10.1103/PhysRevLett.108.082001}{{\em Phys.Rev.Lett.}
  {\bfseries 108} (2012) 082001}
  [\href{http://arxiv.org/abs/1111.1630}{{\ttfamily arXiv:1111.1630}}]
  [\href{http://inspirehep.net/search?p=find+Abelev:2011md}{{\ttfamily
  InSPIRE}}].

\bibitem{Aaij:2013nlm}
{\bfseries LHCb Collaboration} , R.~Aaij {\em et al.}, {\it {Measurement of
  $J/\psi$ polarization in $pp$ collisions at $\sqrt{s}=7$ TeV}},
\href{http://dx.doi.org/10.1140/epjc/s10052-013-2631-3}{{\em Eur.Phys.J.}
  {\bfseries C73} (2013) 2631}
  [\href{http://arxiv.org/abs/1307.6379}{{\ttfamily arXiv:1307.6379}}]
  [\href{http://inspirehep.net/search?p=find+Aaij:2013nlm}{{\ttfamily
  InSPIRE}}].

\bibitem{Chatrchyan:2013cla}
{\bfseries CMS Collaboration} , S.~Chatrchyan {\em et al.}, {\it {Measurement
  of the prompt $J/\psi$ and $\psi(2S)$ polarizations in pp collisions at
  $\sqrt{s}$ = 7 TeV}},
\href{http://dx.doi.org/10.1016/j.physletb.2013.10.055}{{\em Phys.Lett.}
  {\bfseries B727} (2013) 381--402}
  [\href{http://arxiv.org/abs/1307.6070}{{\ttfamily arXiv:1307.6070}}]
  [\href{http://inspirehep.net/search?p=find+Chatrchyan:2013cla}{{\ttfamily
  InSPIRE}}].

\bibitem{Chatrchyan:2012woa}
{\bfseries CMS Collaboration} , S.~Chatrchyan {\em et al.}, {\it {Measurement
  of the Y(1S), Y(2S) and Y(3S) polarizations in $pp$ collisions at
  $\sqrt{s}=7$ TeV}},
\href{http://dx.doi.org/10.1103/PhysRevLett.110.081802}{{\em Phys.Rev.Lett.}
  {\bfseries 110} (2013) 081802}
  [\href{http://arxiv.org/abs/1209.2922}{{\ttfamily arXiv:1209.2922}}]
  [\href{http://inspirehep.net/search?p=find+Chatrchyan:2012woa}{{\ttfamily
  InSPIRE}}].

\bibitem{Gong:2008hk}
B.~Gong and J.-X. Wang, {\it {QCD corrections to polarization of $J/\psi$ and
  $\Upsilon$ at Tevatron and LHC}},
\href{http://dx.doi.org/10.1103/PhysRevD.78.074011}{{\em Phys.Rev.} {\bfseries
  D78} (2008) 074011} [\href{http://arxiv.org/abs/0805.2469}{{\ttfamily
  arXiv:0805.2469}}]
  [\href{http://inspirehep.net/search?p=find+Gong:2008hk}{{\ttfamily
  InSPIRE}}].

\bibitem{Butenschoen:2012px}
M.~Butenschoen and B.~A. Kniehl, {\it {$J/\psi$ polarization at Tevatron and
  LHC: Nonrelativistic-QCD factorization at the crossroads}},
\href{http://dx.doi.org/10.1103/PhysRevLett.108.172002}{{\em Phys.Rev.Lett.}
  {\bfseries 108} (2012) 172002}
  [\href{http://arxiv.org/abs/1201.1872}{{\ttfamily arXiv:1201.1872}}]
  [\href{http://inspirehep.net/search?p=find+Butenschoen:2012px}{{\ttfamily
  InSPIRE}}].

\bibitem{Chao:2012iv}
K.-T. Chao, Y.-Q. Ma, H.-S. Shao, K.~Wang, and Y.-J. Zhang, {\it {$J/\psi$
  Polarization at Hadron Colliders in Nonrelativistic QCD}},
\href{http://dx.doi.org/10.1103/PhysRevLett.108.242004}{{\em Phys.Rev.Lett.}
  {\bfseries 108} (2012) 242004}
  [\href{http://arxiv.org/abs/1201.2675}{{\ttfamily arXiv:1201.2675}}]
  [\href{http://inspirehep.net/search?p=find+Chao:2012iv}{{\ttfamily
  InSPIRE}}].

\bibitem{Gong:2012ug}
B.~Gong, L.-P. Wan, J.-X. Wang, and H.-F. Zhang, {\it {Polarization for Prompt
  $J/\psi$, $\psi(2S)$ production at the Tevatron and LHC}},
\href{http://dx.doi.org/10.1103/PhysRevLett.110.042002}{{\em Phys.Rev.Lett.}
  {\bfseries 110} (2013) 042002}
  [\href{http://arxiv.org/abs/1205.6682}{{\ttfamily arXiv:1205.6682}}]
  [\href{http://inspirehep.net/search?p=find+Gong:2012ug}{{\ttfamily
  InSPIRE}}].

\bibitem{Bodwin:2014gia}
G.~T. Bodwin, H.~S. Chung, U.-R. Kim, and J.~Lee, {\it {Fragmentation
  contributions to $J/\psi$ production at the Tevatron and the LHC}},
\href{http://dx.doi.org/10.1103/PhysRevLett.113.022001}{{\em Phys.Rev.Lett.}
  {\bfseries 113} (2014) 022001}
  [\href{http://arxiv.org/abs/1403.3612}{{\ttfamily arXiv:1403.3612}}]
  [\href{http://inspirehep.net/search?p=find+Bodwin:2014gia}{{\ttfamily
  InSPIRE}}].

\bibitem{Faccioli:2014cqa}
P.~Faccioli, V.~Knunz, C.~Lourenco, J.~Seixas, and H.~K. Wohri, {\it
  {Quarkonium production in the LHC era: a polarized perspective}},
\href{http://dx.doi.org/10.1016/j.physletb.2014.07.006}{{\em Phys.Lett.}
  {\bfseries B736} (2014) 98--109}
  [\href{http://arxiv.org/abs/1403.3970}{{\ttfamily arXiv:1403.3970}}]
  [\href{http://inspirehep.net/search?p=find+Faccioli:2014cqa}{{\ttfamily
  InSPIRE}}].

\bibitem{Gong:2013qka}
B.~Gong, L.-P. Wan, J.-X. Wang, and H.-F. Zhang, {\it {Complete
  next-to-leading-order study on the yield and polarization of
  Upsilon(1S,2S,3S) at the Tevatron and LHC}},
\href{http://dx.doi.org/10.1103/PhysRevLett.112.032001}{{\em Phys.Rev.Lett.}
  {\bfseries 112} (2014) 032001}
  [\href{http://arxiv.org/abs/1305.0748}{{\ttfamily arXiv:1305.0748}}]
  [\href{http://inspirehep.net/search?p=find+Gong:2013qka}{{\ttfamily
  InSPIRE}}].

\bibitem{Han:2014kxa}
H.~Han, Y.-Q. Ma, C.~Meng, H.-S. Shao, Y.-J. Zhang, and K.-T. Chao, {\it
  {$\Upsilon(nS)$ and $\chi_b(nP)$ production at hadron colliders in
  nonrelativistic QCD}},
\href{http://dx.doi.org/10.1103/PhysRevD.94.014028}{{\em Phys. Rev.} {\bfseries
  D94} (2016) 014028} [\href{http://arxiv.org/abs/1410.8537}{{\ttfamily
  arXiv:1410.8537}}]
  [\href{http://inspirehep.net/search?p=find+Han:2014kxa}{{\ttfamily
  InSPIRE}}].

\bibitem{Shao:2014yta}
H.~Shao, H.~Han, Y.~Ma, C.~Meng, Y.~Zhang, {\em et al.},
{\it {Yields and polarizations of prompt $J/\psi$ and $\psi(2S)$ production in
  hadronic collisions}},  [\href{http://arxiv.org/abs/1411.3300}{{\ttfamily
  arXiv:1411.3300}}]
  [\href{http://inspirehep.net/search?p=find+Shao:2014yta}{{\ttfamily
  InSPIRE}}].

\bibitem{Zhang:2009ym}
Y.-J. Zhang, Y.-Q. Ma, K.~Wang, and K.-T. Chao, {\it {QCD radiative correction
  to color-octet $J/\psi$ inclusive production at B Factories}},
\href{http://dx.doi.org/10.1103/PhysRevD.81.034015}{{\em Phys.Rev.} {\bfseries
  D81} (2010) 034015} [\href{http://arxiv.org/abs/0911.2166}{{\ttfamily
  arXiv:0911.2166}}]
  [\href{http://inspirehep.net/search?p=find+Zhang:2009ym}{{\ttfamily
  InSPIRE}}].

\bibitem{Mangano:1996kg}
M.~L. Mangano and A.~Petrelli, {\it {NLO quarkonium production in hadronic
  collisions}},
\href{http://dx.doi.org/10.1142/S0217751X97002048}{{\em Int. J. Mod. Phys.}
  {\bfseries A12} (1997) 3887--3897}
  [\href{http://arxiv.org/abs/hep-ph/9610364}{{\ttfamily hep-ph/9610364}}]
  [\href{http://inspirehep.net/search?p=find+Mangano:1996kg}{{\ttfamily
  InSPIRE}}].

\bibitem{Beneke:1997qw}
M.~Beneke, I.~Z. Rothstein, and M.~B. Wise, {\it {Kinematic enhancement of
  nonperturbative corrections to quarkonium production}},
\href{http://dx.doi.org/10.1016/S0370-2693(97)00832-0}{{\em Phys. Lett.}
  {\bfseries B408} (1997) 373--380}
  [\href{http://arxiv.org/abs/hep-ph/9705286}{{\ttfamily hep-ph/9705286}}]
  [\href{http://inspirehep.net/search?p=find+Beneke:1997qw}{{\ttfamily
  InSPIRE}}].

\bibitem{Fleming:2003gt}
S.~Fleming, A.~K. Leibovich, and T.~Mehen, {\it {Resumming the color octet
  contribution to $e^{+} e^{-} \to J/\psi$ + $X$}},
\href{http://dx.doi.org/10.1103/PhysRevD.68.094011}{{\em Phys.Rev.} {\bfseries
  D68} (2003) 094011} [\href{http://arxiv.org/abs/hep-ph/0306139}{{\ttfamily
  hep-ph/0306139}}]
  [\href{http://inspirehep.net/search?p=find+Fleming:2003gt}{{\ttfamily
  InSPIRE}}].

\bibitem{Fleming:2006cd}
S.~Fleming, A.~K. Leibovich, and T.~Mehen, {\it {Resummation of Large Endpoint
  Corrections to Color-Octet $J/\psi$ Photoproduction}},
\href{http://dx.doi.org/10.1103/PhysRevD.74.114004}{{\em Phys. Rev.} {\bfseries
  D74} (2006) 114004} [\href{http://arxiv.org/abs/hep-ph/0607121}{{\ttfamily
  hep-ph/0607121}}]
  [\href{http://inspirehep.net/search?p=find+Fleming:2006cd}{{\ttfamily
  InSPIRE}}].

\bibitem{Leibovich:2007vr}
A.~K. Leibovich and X.~Liu, {\it {The Color-singlet contribution to $e^{+}
  e^{-} \to J/\psi$ + $X$ at the endpoint}},
\href{http://dx.doi.org/10.1103/PhysRevD.76.034005}{{\em Phys.Rev.} {\bfseries
  D76} (2007) 034005} [\href{http://arxiv.org/abs/0705.3230}{{\ttfamily
  arXiv:0705.3230}}]
  [\href{http://inspirehep.net/search?p=find+Leibovich:2007vr}{{\ttfamily
  InSPIRE}}].

\bibitem{Beneke:1999gq}
M.~Beneke, G.~A. Schuler, and S.~Wolf, {\it {Quarkonium momentum distributions
  in photoproduction and B decay}},
\href{http://dx.doi.org/10.1103/PhysRevD.62.034004}{{\em Phys. Rev.} {\bfseries
  D62} (2000) 034004} [\href{http://arxiv.org/abs/hep-ph/0001062}{{\ttfamily
  hep-ph/0001062}}]
  [\href{http://inspirehep.net/search?p=find+Beneke:1999gq}{{\ttfamily
  InSPIRE}}].

\bibitem{Ma:2015yka}
Y.-Q. Ma, J.-W. Qiu, and H.~Zhang, {\it {Fragmentation functions of polarized
  heavy quarkonium}},
\href{http://dx.doi.org/10.1007/JHEP06(2015)021}{{\em JHEP} {\bfseries 06}
  (2015) 021} [\href{http://arxiv.org/abs/1501.04556}{{\ttfamily
  arXiv:1501.04556}}]
  [\href{http://inspirehep.net/search?p=find+Ma:2015yka}{{\ttfamily InSPIRE}}].

\bibitem{Beneke:1997zp}
M.~Beneke and V.~A. Smirnov, {\it {Asymptotic expansion of Feynman integrals
  near threshold}},
\href{http://dx.doi.org/10.1016/S0550-3213(98)00138-2}{{\em Nucl.Phys.}
  {\bfseries B522} (1998) 321--344}
  [\href{http://arxiv.org/abs/hep-ph/9711391}{{\ttfamily hep-ph/9711391}}]
  [\href{http://inspirehep.net/search?p=find+Beneke:1997zp}{{\ttfamily
  InSPIRE}}].

\bibitem{Collins:2011zzd}
J.~Collins, {\em {Foundations of perturbative QCD}}.
\newblock
\newblock
2011 [\href{http://inspirehep.net/search?p=find+Collins:2011zzd}{{\ttfamily
  InSPIRE}}].
\newblock

\bibitem{Kuhn:1979bb}
J.~H. Kuhn, J.~Kaplan, and E.~G.~O. Safiani, {\it {Electromagnetic Annihilation
  of $e^+e^-$ Into Quarkonium States with Even Charge Conjugation}},
\href{http://dx.doi.org/10.1016/0550-3213(79)90055-5}{{\em Nucl.Phys.}
  {\bfseries B157} (1979) 125}
  [\href{http://inspirehep.net/search?p=find+Kuhn:1979bb}{{\ttfamily
  InSPIRE}}].

\bibitem{Guberina:1980dc}
B.~Guberina, J.~H. Kuhn, R.~Peccei, and R.~Ruckl, {\it {Rare Decays of the
  $Z^0$}},
\href{http://dx.doi.org/10.1016/0550-3213(80)90287-4}{{\em Nucl.Phys.}
  {\bfseries B174} (1980) 317}
  [\href{http://inspirehep.net/search?p=find+Guberina:1980dc}{{\ttfamily
  InSPIRE}}].

\bibitem{Fan:2009cj}
Y.~Fan, Z.-G. He, Y.-Q. Ma, and K.-T. Chao, {\it {Predictions of Light Hadronic
  Decays of Heavy Quarkonium $^1D_2$ States in NRQCD}},
\href{http://dx.doi.org/10.1103/PhysRevD.80.014001}{{\em Phys.Rev.} {\bfseries
  D80} (2009) 014001} [\href{http://arxiv.org/abs/0903.4572}{{\ttfamily
  arXiv:0903.4572}}]
  [\href{http://inspirehep.net/search?p=find+Fan:2009cj}{{\ttfamily InSPIRE}}].

\bibitem{Kang:2014pya}
Z.-B. Kang, Y.-Q. Ma, J.-W. Qiu, and G.~Sterman, {\it {Heavy Quarkonium
  Production at Collider Energies: Partonic Cross Section and Polarization}},
\href{http://dx.doi.org/10.1103/PhysRevD.91.014030}{{\em Phys.Rev.} {\bfseries
  D91} (2015) 014030} [\href{http://arxiv.org/abs/1411.2456}{{\ttfamily
  arXiv:1411.2456}}]
  [\href{http://inspirehep.net/search?p=find+Kang:2014pya}{{\ttfamily
  InSPIRE}}].

\bibitem{Braaten:1993mp}
E.~Braaten, K.~Cheung, and T.~C. Yuan, {\it {$Z^0$ decay into charmonium via
  charm quark fragmentation}},
\href{http://dx.doi.org/10.1103/PhysRevD.48.4230}{{\em Phys.Rev.} {\bfseries
  D48} (1993) 4230--4235}
  [\href{http://arxiv.org/abs/hep-ph/9302307}{{\ttfamily hep-ph/9302307}}]
  [\href{http://inspirehep.net/search?p=find+Braaten:1993mp}{{\ttfamily
  InSPIRE}}].

\bibitem{Braaten:1993rw}
E.~Braaten and T.~C. Yuan, {\it {Gluon fragmentation into heavy quarkonium}},
\href{http://dx.doi.org/10.1103/PhysRevLett.71.1673}{{\em Phys.Rev.Lett.}
  {\bfseries 71} (1993) 1673--1676}
  [\href{http://arxiv.org/abs/hep-ph/9303205}{{\ttfamily hep-ph/9303205}}]
  [\href{http://inspirehep.net/search?p=find+Braaten:1993rw}{{\ttfamily
  InSPIRE}}].

\bibitem{Braaten:1994kd}
E.~Braaten and T.~C. Yuan, {\it {Gluon fragmentation into P wave heavy
  quarkonium}},
\href{http://dx.doi.org/10.1103/PhysRevD.50.3176}{{\em Phys.Rev.} {\bfseries
  D50} (1994) 3176--3180}
  [\href{http://arxiv.org/abs/hep-ph/9403401}{{\ttfamily hep-ph/9403401}}]
  [\href{http://inspirehep.net/search?p=find+Braaten:1994kd}{{\ttfamily
  InSPIRE}}].

\bibitem{Braaten:1995cj}
E.~Braaten and T.~C. Yuan, {\it {Gluon fragmentation into spin triplet S wave
  quarkonium}},
\href{http://dx.doi.org/10.1103/PhysRevD.52.6627}{{\em Phys. Rev.} {\bfseries
  D52} (1995) 6627--6629}
  [\href{http://arxiv.org/abs/hep-ph/9507398}{{\ttfamily hep-ph/9507398}}]
  [\href{http://inspirehep.net/search?p=find+Braaten:1995cj}{{\ttfamily
  InSPIRE}}].

\bibitem{Ma:1995vi}
J.~P. Ma, {\it {Quark fragmentation into p wave triplet quarkonium}},
\href{http://dx.doi.org/10.1103/PhysRevD.53.1185}{{\em Phys. Rev.} {\bfseries
  D53} (1996) 1185--1190}
  [\href{http://arxiv.org/abs/hep-ph/9504263}{{\ttfamily hep-ph/9504263}}]
  [\href{http://inspirehep.net/search?p=find+Ma:1995vi}{{\ttfamily InSPIRE}}].

\bibitem{Braaten:2000pc}
E.~Braaten and J.~Lee, {\it {Next-to-leading order calculation of the color
  octet 3S(1) gluon fragmentation function for heavy quarkonium}},
\href{http://dx.doi.org/10.1016/S0550-3213(00)00396-5}{{\em Nucl. Phys.}
  {\bfseries B586} (2000) 427--439}
  [\href{http://arxiv.org/abs/hep-ph/0004228}{{\ttfamily hep-ph/0004228}}]
  [\href{http://inspirehep.net/search?p=find+Braaten:2000pc}{{\ttfamily
  InSPIRE}}].

\bibitem{Bodwin:2003wh}
G.~T. Bodwin and J.~Lee, {\it {Relativistic corrections to gluon fragmentation
  into spin triplet S wave quarkonium}},
\href{http://dx.doi.org/10.1103/PhysRevD.69.054003}{{\em Phys. Rev.} {\bfseries
  D69} (2004) 054003} [\href{http://arxiv.org/abs/hep-ph/0308016}{{\ttfamily
  hep-ph/0308016}}]
  [\href{http://inspirehep.net/search?p=find+Bodwin:2003wh}{{\ttfamily
  InSPIRE}}].

\bibitem{Hao:2009fa}
G.~Hao, Y.~Zuo, and C.-F. Qiao,
{\it {The Fragmentation Function of Gluon Splitting into P-wave Spin-singlet
  Heavy Quarkonium}},  [\href{http://arxiv.org/abs/0911.5539}{{\ttfamily
  arXiv:0911.5539}}]
  [\href{http://inspirehep.net/search?p=find+Hao:2009fa}{{\ttfamily InSPIRE}}].

\bibitem{Jia:2012qx}
Y.~Jia, W.-L. Sang, and J.~Xu, {\it {Inclusive $h_c$ Production at $B$
  Factories}},
\href{http://dx.doi.org/10.1103/PhysRevD.86.074023}{{\em Phys. Rev.} {\bfseries
  D86} (2012) 074023} [\href{http://arxiv.org/abs/1206.5785}{{\ttfamily
  arXiv:1206.5785}}]
  [\href{http://inspirehep.net/search?p=find+Jia:2012qx}{{\ttfamily InSPIRE}}].

\bibitem{Bodwin:2012xc}
G.~T. Bodwin, U.-R. Kim, and J.~Lee, {\it {Higher-order relativistic
  corrections to gluon fragmentation into spin-triplet S-wave quarkonium}},
\href{http://dx.doi.org/10.1007/JHEP11(2012)020}{{\em JHEP} {\bfseries 1211}
  (2012) 020} [\href{http://arxiv.org/abs/1208.5301}{{\ttfamily
  arXiv:1208.5301}}]
  [\href{http://inspirehep.net/search?p=find+Bodwin:2012xc}{{\ttfamily
  InSPIRE}}].

\bibitem{Ma:2013yla}
Y.-Q. Ma, J.-W. Qiu, and H.~Zhang, {\it {Heavy quarkonium fragmentation
  functions from a heavy quark pair. I. $S$ wave}},
\href{http://dx.doi.org/10.1103/PhysRevD.89.094029}{{\em Phys.Rev.} {\bfseries
  D89} (2014) 094029} [\href{http://arxiv.org/abs/1311.7078}{{\ttfamily
  arXiv:1311.7078}}]
  [\href{http://inspirehep.net/search?p=find+Ma:2013yla}{{\ttfamily InSPIRE}}].

\bibitem{Ma:2014eja}
Y.-Q. Ma, J.-W. Qiu, and H.~Zhang, {\it {Heavy quarkonium fragmentation
  functions from a heavy quark pair. II. $P$ wave}},
\href{http://dx.doi.org/10.1103/PhysRevD.89.094030}{{\em Phys.Rev.} {\bfseries
  D89} (2014) 094030} [\href{http://arxiv.org/abs/1401.0524}{{\ttfamily
  arXiv:1401.0524}}]
  [\href{http://inspirehep.net/search?p=find+Ma:2014eja}{{\ttfamily InSPIRE}}].

\bibitem{Bodwin:2015iua}
G.~T. Bodwin, K.-T. Chao, H.~S. Chung, U.-R. Kim, J.~Lee, and Y.-Q. Ma, {\it
  {Fragmentation contributions to hadroproduction of prompt $J/\psi$ ,
  $\chi_{cJ}$ , and $\psi(2S)$ states}},
\href{http://dx.doi.org/10.1103/PhysRevD.93.034041}{{\em Phys. Rev.} {\bfseries
  D93} (2016) 034041} [\href{http://arxiv.org/abs/1509.07904}{{\ttfamily
  arXiv:1509.07904}}]
  [\href{http://inspirehep.net/search?p=find+Bodwin:2015iua}{{\ttfamily
  InSPIRE}}].

\bibitem{machao}
Y.-Q. Ma and K.-T. Chao {\em In preparation} .

\bibitem{Xu:2012am}
G.-Z. Xu, Y.-J. Li, K.-Y. Liu, and Y.-J. Zhang, {\it {Relativistic Correction
  to Color Octet $J/\psi$ Production at Hadron Colliders}},
\href{http://dx.doi.org/10.1103/PhysRevD.86.094017}{{\em Phys.Rev.} {\bfseries
  D86} (2012) 094017} [\href{http://arxiv.org/abs/1203.0207}{{\ttfamily
  arXiv:1203.0207}}]
  [\href{http://inspirehep.net/search?p=find+Xu:2012am}{{\ttfamily InSPIRE}}].

\end{thebibliography}
\end{document}